\documentclass[preprint,number,12pt]{elsarticle}

\usepackage{graphicx}
\usepackage{mathptmx}      
\usepackage{amssymb}
\biboptions{comma}
 
\journal{Astroparticle Physics}

\begin{document} 

\begin{frontmatter}

\title{Prospects for Observations of Pulsars and Pulsar Wind Nebulae with CTA}

\author[mpik,ICE]{E. de O\~na-Wilhelmi\corref{cor1}} \ead{emma@mpi-hd.mpg.de}
\author[PAN,PAN2]{B. Rudak\corref{cor1}} \ead{bronek@ncac.torun.pl}
\author[UCM]{J. A. Barrio}
\author[UCM]{J. L. Contreras}
\author[Mont]{Y. Gallant}
\author[ICE]{D. Hadasch}
\author[UCM]{T. Hassan}
\author[UCM]{M. Lopez}
\author[IFAE]{D. Mazin}
\author[UCM]{N. Mirabal}
\author[ICE]{G. Pedaletti}
\author[Mont]{M. Renaud}
\author[mpik]{R. de los Reyes}
\author[ICE,ICREA]{D. F. Torres}
\author[]{for the CTA collaboration}
\cortext[cor1]{Corresponding author}

\address[mpik]{Max-Planck-Institut f\"ur Kernphysik, P.O. Box 103980, D 69029 Heidelberg, Germany}
\address[PAN]{Centrum Astronomiczne im. M. Kopernika PAN, Bartycka 18, 00-716 Warszawa, Poland}
\address[PAN2]{Katedra Astronomii i Astrofizyki CA, UMK Torun, Poland}
\address[UCM]{Dpto. de F\'isica At\'omica, Molecular y
Nuclear, Universidad Complutense de 
Madrid, Spain}
\address[Mont]{Laboratoire Univers et Particules de Montpellier, 
Universit\'e Montpellier 2, CNRS/IN2P3, CC 072, 
Place Eug\`ene Bataillon, F-34095 Montpellier Cedex 5, France}
\address[ICE]{Institut de Ci\`encies de l'Espai (IEEC-CSIC),
             Campus UAB,  Torre C5, 2a planta,
             08193 Barcelona, Spain}
\address[IFAE]{IFAE, Edifici Cn. Universitat Autonoma de Barcelona, Barcelona, Spain}
\address[ICREA]{Instituci\'o Catalana de Recerca i Estudis Avan\c{c}ats (ICREA)}

\begin{abstract}
The last few years have seen a revolution in very-high $\gamma$-ray astronomy (VHE; E$>$100 GeV) driven largely by a new generation of Cherenkov telescopes (namely the H.E.S.S. telescope array, the MAGIC and MAGIC-II large telescopes and the VERITAS telescope array). The Cherenkov Telescope Array (CTA) project foresees a factor of 5 to 10 improvement in sensitivity above 0.1 TeV, extending the accessible energy range to higher energies up to 100 TeV, in the Galactic cut-off regime, and down to a few tens GeV, covering the VHE photon spectrum with good energy and angular resolution. 
As a result of the fast development of the VHE field, the number of pulsar wind nebulae (PWNe) detected has increased from one PWN in the early '90s to more than two dozen firm candidates today. Also, the low energy threshold achieved and good sensitivity at TeV energies has resulted in the detection of pulsed emission from the Crab Pulsar (or its close environment) opening new and exiting expectations about the pulsed spectra of the high energy pulsars powering PWNe. Here we discuss the physics goals we aim to achieve with CTA on pulsar and PWNe physics evaluating the response of the instrument for different configurations.

\end{abstract}

\begin{keyword}
\end{keyword}

\end{frontmatter}

\section{Pulsars and Pulsar Wind Nebulae at Very High Energy}
\label{sec:1}
Pulsars and their synchrotron nebula have been extensively observed in radio and X-ray wavelengths. The non-thermal synchrotron emission detected from these systems provides evidence for particle (lepton) acceleration to extremely high energies, susceptible to inverse-Compton (IC) scattering to target soft photons in the surrounding medium and produce VHE $\gamma$-ray emission.
Recent observations of Galactic sources with Imaging Cherenkov Atmospheric telescopes (IACTs) have revealed PWNe to be the most effective Galactic objects for the production of VHE $\gamma$-ray emission. As recently as 2004, only the Crab PWN was detected with a steady $\gamma$-ray flux above 1\,TeV of (2.1$\pm$0.1$_{\rm stat}$)$\times$10$^{\rm -11}$cm$^{\rm -2}$s$^{-1}$ \cite{HEGRACrab,whipplecrab}. The development of new sensitive IACTs in the last years has raised the number of likely PWNe detected to at least 27 sources, whereas many of the unidentified $\gamma$-ray sources are widely believed to be PWNe (or old relic PWNe).

\begin{figure}[t]
\includegraphics[scale=.6]{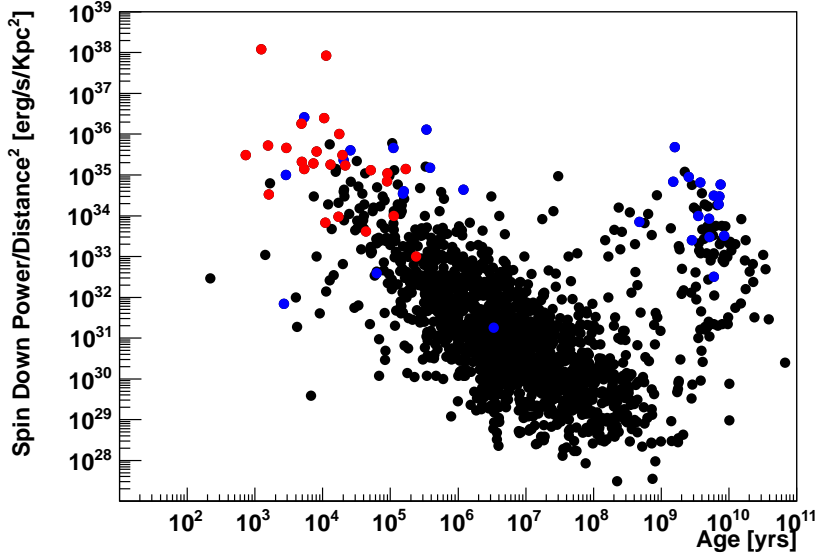}
%
%
\caption{Spin-down energy loss rate of radio pulsars listed in the ATNF catalog versus their characteristic age. The blue dots overlaid indicate the pulsars detected at high energies with the Fermi-LAT telescope whereas the red ones mark the pulsars associated to TeV PWNe.}
\label{fig1}       
\end{figure}

These VHE PWNe are believed to be related to the young and energetic pulsars that power highly magnetized nebulae (a few $\mu$G to a few hundred $\mu$G). In this scenario, particles are accelerated to VHE along
their expansion into the pulsar surroundings, or at the shocks produced in
collisions of the winds with the surrounding medium. Leptons accelerated to relativistic speeds interact with the low energy radiations fields produced by synchrotron, thermal or microwave-background origins. As a result, non-thermal radiation is produced from the lowest
possible energies up to $\sim$100 TeV. For a few $\mu$G magnetic fields, freshly injected electrons (and positrons) create a small synchrotron
nebula around the pulsar which should be visible in X-rays, in
contrast to an often much larger TeV nebula, generated by IC processes (for a recent review see \cite{pwn}). 
Typically only young pulsars with large spin-down power ($\dot{E}$$>$10$^{\rm 33}$erg\,s$^{\rm -1}$) produce prominent PWNe \cite{HighPulsars}. This is illustrated in Fig. \ref{fig1} where the spin-down power of radio pulsars from the ATNF catalog \cite{ATNF} is shown versus their characteristic age. Pulsars powering VHE $\gamma$-ray PWNe are marked with red bullets whereas pulsars in which periodic emission originating in the pulsar magnetosphere (as detected by the Fermi-LAT satellite, 100 MeV$<$E$<$100 GeV) are shown in blue. 

The discovery of this large population of TeV PWNe has provided a large input for multi-wavelength models of particle evolution and acceleration in these objects. Theoretical MHD models (carried out by e.g. \cite{Zhang2008}) describe well the observed emission by a relativistic wind of particles driven by a central pulsar blowing into the ambient medium creating a termination shock. The particles are believed to be accelerated somewhere between the light-cylinder and the termination shock, although this question is still open. The broadband spectrum of a PWN thus provides constraints on the integrated energy injected by the pulsar as well as on the effects of adiabatic expansion and the evolution of the magnetic field. The spectral energy distribution (SED) consists of two components, 1) synchrotron emission extending from the radio into the X-ray and, in some cases, the MeV band, and 2) IC emission producing GeV and TeV photons. 
The combined observations at VHE, X-rays and radio wavelength are crucial to constrain the evolution of the nebula magnetic field as well as the magnetic-to-kinetic energy conversion \cite{dd09}.

Pulsars are sources of non-thermal radiation which in some cases
extends over many decades in energy, from radio bands to $\gamma$-rays.
Thanks to Fermi LAT observations pulsars are now acknowledged as a distinct class of galactic $\gamma$-ray sources.
Fermi LAT Second Source Catalog \cite{2fgl} contains nearly 90 individual pulsars
including 27 firm detections of millisecond pulsars (MSPs).
Different models of pulsar activity give different predictions for emission characteristics
in terms of directionality and energy spectra. 
Due to the highly anisotropic (beamed) and energy-dependent structure of  
pulsar emission the observed radiation patterns, i.e.\ the light curves and spectra, depend strongly
on the line-of-sight with respect to the spin axis.
In most cases, a simple power-law with exponential cutoff located at a few GeV fits the LAT spectra satisfactorily.
The $\gamma$-ray light curves are mostly either single-peaked or double-peaked; their absolute phasing with radio light curves provide important clues to determine topology and spatial extent of the emission region.
It came as a surprise that in most cases of MSPs their $\gamma$-ray properties are similar to those of normal pulsars despite much weaker magnetic field strengths of the former. 
Moreover, unpulsed $\gamma$-rays from 11 globular clusters have been detected; this is interpreted as
due to the ensembles of MSPs in these clusters. Some support for such interpretation comes from the case 
of J1823--3021A in globular cluster NGC6624,
the first MSP in a GC detected so far
as a source of pulsed $\gamma$-rays \cite{webb}. 

Little is known, however, about the properties of pulsars above $\approx$10 GeV where LAT performance is limited. Recent results from VERITAS \cite{veritas} and MAGIC \cite{mono,stereo} have shown, for the Crab pulsar, a clear detection, up to 400 GeV of pulsed power-law type emission, as opposed to an exponential cutoff as might have been expected. 

\subsection{PWNe physics goals with CTA}
Observations at VHE have revealed PWNe as the most efficient type of source in accelerating particles and producing $\gamma$-ray through IC mechanisms, even allowing the detection of such systems outside our own Galaxy \cite{lmc}. Moreover, the VHE part of the SED or IC peak contains information of the total energy released by the pulsar in its life time. This is due to the slower cooling time of TeV electrons ($\tau\approx$(4.8 kyr)B$^{-2}_{-5}$E$^{-1/2}_{\rm TeV}$) compared with the live-time of keV-emitting electrons ($\tau\approx$(1.2 kyr)B$^{-3/2}_{-5}$E$^{-1/2}_{\rm keV}$). PWNe thus act like calorimeters at VHE and observations with Cherenkov telescopes, and in particular with CTA, permit the investigation of the time-evolution of such systems. This also implies a better understanding of the still unknown mechanism of how the pulsar releases its rotational energy from a cold relativistic wind blowing out the light cylinder to a kinetic one with Lorenz factor of $\approx$3$\times$10$^{6}$ at the termination shock. Within the context of temporal evolution of PWNe, two types of object seem to be emerging: 1) young plerions such as the Crab nebula \cite{HEGRACrab}, SNR\,G0.9+0.1 \cite{G09}, MSH\,15--52 \cite{MSH} and the newly discovered Crab-like VHE $\gamma$-ray sources SNR\,G21.5--0.9 and Kes\,75 \cite{kes75} and 2) evolved (extended and resolved) systems (i.e. with characteristic ages $\tau$$>$10$^{4}$ yr), such as Vela X \cite{velax}, HESS\,J1825--137 \cite{1825}, HESS\,J1718--385 \cite{1718} and HESS\,J1809--193 \cite{1809}.

The young systems show in general a good match with the morphology seen in X-rays, while in the latter group, very often the pulsar powering the TeV plerion is found offset with respect to the center of the TeV emission, with large size ratios between the X-ray and VHE $\gamma$-ray emission regions. The evolution of the supernova remnant (SNR) blast wave into an inhomogeneous ISM and/or the high velocity of the pulsar, together with a low magnetic field value ( $\approx $5$\mu$G), may explain these large offsets as being the relic nebulae from the past history of the pulsar wind inside its host SNR. Electrons responsible of the VHE radiation (with energy E$_{\rm e}$=(18\,TeV)E$^{1/2}_{\rm TeV}$ to inverse Compton scatter CMBR seed photons to energies E$_{\rm TeV}$, to be compared with the required electron energy in a transverse magnetic field of strength B = 10$^{-5}$B$_{-5}$\,G, to radiate synchrotron photons of mean energy E$_{\rm keV}$,  E$_{\rm e}$=(70\,TeV)B$^{-1/2}_{-5}$E$^{1/2}_{\rm keV}$) do not suffer from severe radiative losses and the majority of them may survive from (and hence probe) early epochs of the PWNe evolution. This interpretation has been further supported by the discovery of the spectral softening of the VHE nebula HESS\,J1825--137 as a function of the distance from the pulsar \cite{1825}. 
All these new observations have allowed a fast development of the field and a better understanding of the physics of PWNe as well as opened new questions about e.g. particle transport and dynamics, energy loses, inter-play between the high spin-down energy pulsar and its surrounding nebula and interaction between PWNe and the hosting SNR. 

As the largest and most efficient population of VHE $\gamma$-rays sources, it is clear that a deeper understanding of PWNe is one of the key goals of the CTA observatory. The following research lines have been pursued and evaluated for the three different configurations described previously in this volume to asses the CTA best response:

\begin{itemize}
\item {\bf Maximum and minimum resolvable size of the PWN.}

The majority of VHE PWNe have a very large size up to 1.2$^{\rm o}$ depending on their evolutionary stage and proximity (see e.g. Vela\,X or HESS\,J1825--137). One of the key factors for the exponential increase of the PWN number in the last years has been the large field of view of the third generation of Cherenkov Telescopes ($\approx$5$^{\rm o}$), which has allowed the detection of the large $\gamma$-ray halo produced by cooled electrons. Similar features will be required for CTA to be able to image the whole plerionic emission. At least a flat response up to 1.5$^{\rm o}$ should be aimed for. 

Together with a large field of view, to be able to resolve sub-structures on the VHE emission and thus sites of VHE $\gamma$-ray production, a good angular resolution is required of at least $<$3$^\prime$. Given the smaller size and complexity of the nebulae detected in X-rays, an angular resolution comparable to systematic pointing error (5$^{\prime\prime}$ per axis) should be achieved for CTA (as well for the study of composite SNRs, see below). 

\item{\bf Electron cooling effect}
The large energy range covered by CTA will be crucial to understand cooling effects in the electron parent populations, including the resolution of internal structures and the possibility to disentangle between synchrotron or adiabatic losses. 

\item{\bf Composite SNRs}
One of the most intriguing aspects of the new observations is the interaction between the host SNR and its PWN, and how much of the $\gamma$-ray emission is originated by the SNR. A good angular resolution is required to understand those composed systems, which often are unresolved when observed with the present instruments (i.e. Kes\,75 with a shell radius of 1.75$^\prime$ or SNR\,G21.5--0.9, with a radius of 2.5$^\prime$, to compare with the current Cherenkov telescopes angular resolution of $\approx$3$^\prime$). 

\item{\bf Population studies}
PWNe have been proven to be the most efficient Galactic sources in producing VHE $\gamma$-ray. This large efficiencies and large size (detection sensitivity is proportional to 1/d, where d is the source size, for extended sources and not 1/d$^2$) permit the detection of PWNe up to 50 kpc (i.e. PSR\,J0537--6910 on the LMC). Observations with CTA will allow an homogeneous sampling of the Galaxy. 

\item{\bf PWN Modeling}
The new observations and results will constrain and provide input for MHD simulations of PWNe, such as estimation of magnetic field, electron population behind the $\gamma$-ray (and synchrotron) emission, spectral characteristics, etc . \end{itemize}

\subsection{Pulsars physics goals with CTA}
One of the key issues in pulsar studies is the question on how far the non-thermal spectra of pulsed radiation
from known $\gamma$-ray pulsars
actually extend in the energy domain and what are the shapes of their cutoffs.
The shape and position of the spectral cutoff vary depending on the emission model considered and the observer orientation.  In polar cap models 
(e.g. \cite{dh96}) electrons are accelerated at low altitudes and radiate $\gamma$-rays via synchrotron-curvature radiation. Since these $\gamma$-rays are created in super-strong magnetic fields, magnetic pair production is unavoidable, and hence, only those secondary photons which survive pair creation (a few GeV for typical pulsar magnetosphere) escape to infinity as an observed pulsed emission. A natural consequence of the polar cap process is a super-exponential cutoff of the spectrum above a characteristic energy E$_{\rm o}$. 
The way to maintain the magnetic pair creation above the gap and avoid super-exponential cutoffs is to extend the gaps
spatially. This is a property of the slot-gap model by \cite{MH2004}. In outer gap models, e.g. \cite{hk08}, $\gamma$-ray production is expected to occur further away from the stellar surface. 
In this case the cutoff is determined either by photon-photon pair production
or by maximal energy reached via inverse Compton scattering.
Apparent absence of super-exponential cutoffs in the Fermi LAT pulsar spectra
have called into question, therefore, the classical picture of polar cap activity in pulsars.

One of the main motivations of CTA is to explore the energy range of 10 GeV to 100 GeV, 
by using very large reflectors 
to decrease the energy threshold to a few tens of GeV.
CTA should thus offer a unique opportunity 
to explore the most extreme energetic processes in
pulsar magnetospheres and beyond their light cylinders. 
The goal is to formulate
self-consistent electrodynamic theory of pulsar activity and the VHE domain 
should play an important role in it.

\subsubsection{Pulsed VHE from the Crab pulsar - the theoretical context}
Pulsed VHE emission from the Crab pulsar is of particular interest for pulsar physics.
In the context of outer gap models its presence and properties depend strongly on the actual location
of outer boundary of the outer gap with respect to the light cylinder. 
Above a few GeV it is
synchro-curvature emission due to primary electrons which dominates all other intrinsic spectral components \cite{hk08,hk10}.
However, it is strongly attenuated by the dense field of magnetospheric soft photons.
The emerging radiation is then dominated by IC scattering of secondary and tertiary pairs with the soft photon field.

There have been other suggestions about possible sources of pulsed HE and/or VHE radiation
from pulsars, like the slot-gap model (\cite{MH2004}), 
the magnetospheres with a force-free structure \cite{bs10}, the striped wind model \cite{petri} or IC scattering of soft thermal and non-thermal photons  
from the pulsar with ultra-relativistic electrons in its pulsar wind zone \cite{ab, AHA2012}.

In all present-day outer-gap models the location
of the outer boundary (as well as the inner boundary) of the outer gap is not determined self-consistently;
therefore, it is a free parameter. The particular choice of the outer boundary location
results in a particular extent and shape of the HE-VHE radiation:
The outer boundary located at around 75\% of the light cylinder radius
results in the spectrum with exponential cutoff before even reaching 25 GeV \cite{hk08}.
But the outer boundary closer to the light cylinder leads to a lower attenuation probability
of the intrinsic VHE photons in the field of soft photons.
In particular, the VHE spectral component in pulsed radiation from the Crab pulsar
as discovered by VERITAS \cite{veritas} and MAGIC \cite{mono,stereo} 
may be accommodated by shifting the outer boundary to a higher value (K. Hirotani, private communication); moreover, the cutoff shape changes 
from an exponential-like to a double-power-law (Fig.15 in \cite{mono}).
This does not mean,  however, that a power law combined with an exponential cutoff to describe the
pulsar spectra as inferred from Fermi LAT are invalidated with the VERITAS discovery.
The power law with exponential cutoff
is just one of several template shapes used in the spectral analysis of Fermi LAT pulsar data.  
A statement that the best fit for a given pulsar data is a power law with exponential cutoff does not
rule out significant deviations from this shape, especially at the highest energy bins.
With large error bars at the highest energy bins there may be plenty
of room to accommodate an additional component (e. g. expected due to some ICS at VHE).

The discovery of pulsed VHE emission shows that the outer gap models should undergo major
modifications if the gaps are to be closer to the light cylinder than assumed before.
More accurate treatment of
the electrodynamics, including currents and possibly going beyond the light cylinder
is necessary. High quality spectral, phase-resolved properties of the VHE radiation
from Crab will be essential to help to develop realistic models of the magnetospheric gaps.
CTA is expected, therefore, to contribute significantly
to the progress in our understanding of the magnetospheric activity of high-spin-down pulsars.

\subsubsection{Unpulsed VHE from Globular Clusters}
Globular Clusters are known to contain many MSPs and they are expected to be the sources of unpulsed VHE radiation e.g. \cite{bs,ven}. The most
attractive globular cluster harboring probably as many as 200\,MSPs (more than 20 radio-pulsars are known to date) is GC 47\,Tucanae (47\,Tuc).
The upper limit from H.E.S.S. on the VHE flux above 800\,GeV from 47\,Tuc is at the level 6.7$\times$10$^{-13}$cm$^{-2}$s$^{-1}$ \cite{tuc47}.
The unpulsed VHE flux expected from 47\,Tuc in the model by \cite{ven} is the sum of two IC scattering components due to primary
electrons up-scattering ambient starlight and CMB photons but its level depends heavily on the strength of ambient magnetic field in the cluster. 
The H.E.S.S. UL appears stringent enough to constrain the parameter space (number of MSPs vs. ambient magnetic field strength) in
the simple model of 47\,Tuc \cite{ven}.
H.E.S.S. reported recently on the discovery of the VHE radiation from vicinity of Terzan\,5 (Ter\,5) \cite{ter5}. This radiation
may be due to MSPs in the cluster, however a significant offset from the location of Ter\,5 has no satisfactory explanation so far.
For other globular clusters, with lower content of MSPs, the expected flux level between 100\,GeV and 10\,TeV may not be within the reach of present-day IACTs. However, CTA may be sensitive enough to detect some globular clusters
and even to resolve their spectral shapes at VHE.\\
CTA will, therefore, in principle 
allow the detailed study of the major ingredients of globular clusters which link MSPs to the expected VHE emission.\\

\section{CTA response simulations}
\subsection{Pulsar Wind Nebulae}
\subsubsection{The Crab Nebula}
The Crab Nebula (M1, NGC\,1952) is the first source detected at TeV energies and it is considered traditionally the standard candle of the TeV sky. It was first detected in 1989 with the Whipple Cherenkov telescope \citep{whipplecrab} followed by the HEGRA telescopes \citep{HEGRACrab} which reported and confirmed a bright steady source (at the instrument's sensitivity)  with a flux at 1\,TeV of (2.83$\pm$0.04)$\cdot$10$^{-11}$ TeV$^{-1}$s$^{-1}$cm$^{-2}$ and a spectral index of 2.62$\pm$0.02. With the third generation of Cherenkov telescopes, namely MAGIC, H.E.S.S. and VERITAS, the Crab Nebula has been extensively observed, putting constraints to the low and high ends of the energy spectrum, size and variability \citep{magiccrab,hesscrab,fermiflares,agileflares}. But even if the Crab Nebula is perhaps the best studied/observed source in the TeV sky, it is still one of the most mysterious sources. The multi-wavelength nebula spectrum shows an undisputed synchrotron nature of the non-thermal radiation from radio to low energy $\gamma$-rays, indicating the existence of relativistic electrons of energies up to 10$^{\rm 16}$ eV. The Compton scattering of the same electrons leads to effective VHE emission while the synchrotron component is responsible for the radiation from radio to relatively low energy $\gamma$-rays (E$<$1\,GeV) \cite{CrabModel}. Recently, the AGILE and Fermi collaborations have reported flaring activity above 1\,GeV from the nebula \citep{fermiflares,agileflares}. If these flares are related to synchrotron emission of electrons from the wind termination shocks, only the highest energy electrons above 100\,TeV (in fact much higher than 100\,TeV)  can be responsible for this emission in the E$>$100 MeV band and only these
electrons can provide changes on scales of days for any reasonable assumption about the magnetic field. That implies a corresponding IC photon flux above at least 10\,TeV. 

Observations with CTA will be crucial to finally understand the non-thermal processes occurring within the Crab nebula. At high energies, CTA should be able to constrain the maximum energy of the accelerated electrons and determine the cutoff energy with precision. Searches of variation of the spectrum at the highest energies should help us to unveil the origin of the Fermi and AGILE detected flares at short time scales. These tails may be also visible at the lowest part of the CTA-detected energy spectrum if the energy threshold is low enough. 

We evaluated three proposed configurations, CTA\_B\_ifae, CTA\_C\_ifae and  CTA\_E\_ifae \footnote{Monte Carlo Simulation from IFAE using configuration B -- optimized for low energy events, C -- optimized for high energy events and E -- balanced configuration between the lowest and highest energies} (B, C, E) using {\it CTAmacrosv6} for two different energy spectra. For the lowest energies, we used the Crab Nebula spectrum measured by the MAGIC telescope. The photon spectrum between 60\,GeV and 9\,TeV is well described by a curved power-law dN/dE=I$_{\rm o}$(E/300\,GeV)$^{\rm [-a-b\cdot log10(E/300 GeV)]}$ with a flux normalization at 300\,GeV of I$_{\rm o}$=(6.0 $\pm$ 0.2)$\cdot$10$^{-10}$TeV$^{-1}$s$^{-1}$cm$^{-2}$, a=-2.31$\pm$0.06 and b=-0.26$\pm$0.07\,TeV. To evaluate the cutoff energy, we used the spectrum measured by H.E.S.S. between 440\,GeV up to 20\,TeV. A fit to a power-law function with an exponential cutoff dN/dE=I$_{\rm o}$(E/1 TeV)$^{-\Gamma}$exp(-E/E$_{\rm cut}$) results on a differential flux normalization at 1\,TeV of (3.45$\pm$0.07)$\cdot$10$^{-11}$TeV$^{-1}$s$^{-1}$cm$^{-2}$, with $\Gamma$=2.39$\pm$0.03 and a cutoff energy E$_{\rm cut}$=14.3$\pm$2.1\,TeV. These two spectral shape are simulated and the reconstructed spectra as detected by CTA with the different configurations are fitted with a MAGIC- and HESS-like function. The results of the fits are listed in Table \ref{table:1}. 

\begin{figure}[t!]
 \centering
 \includegraphics[width=0.45\textwidth]{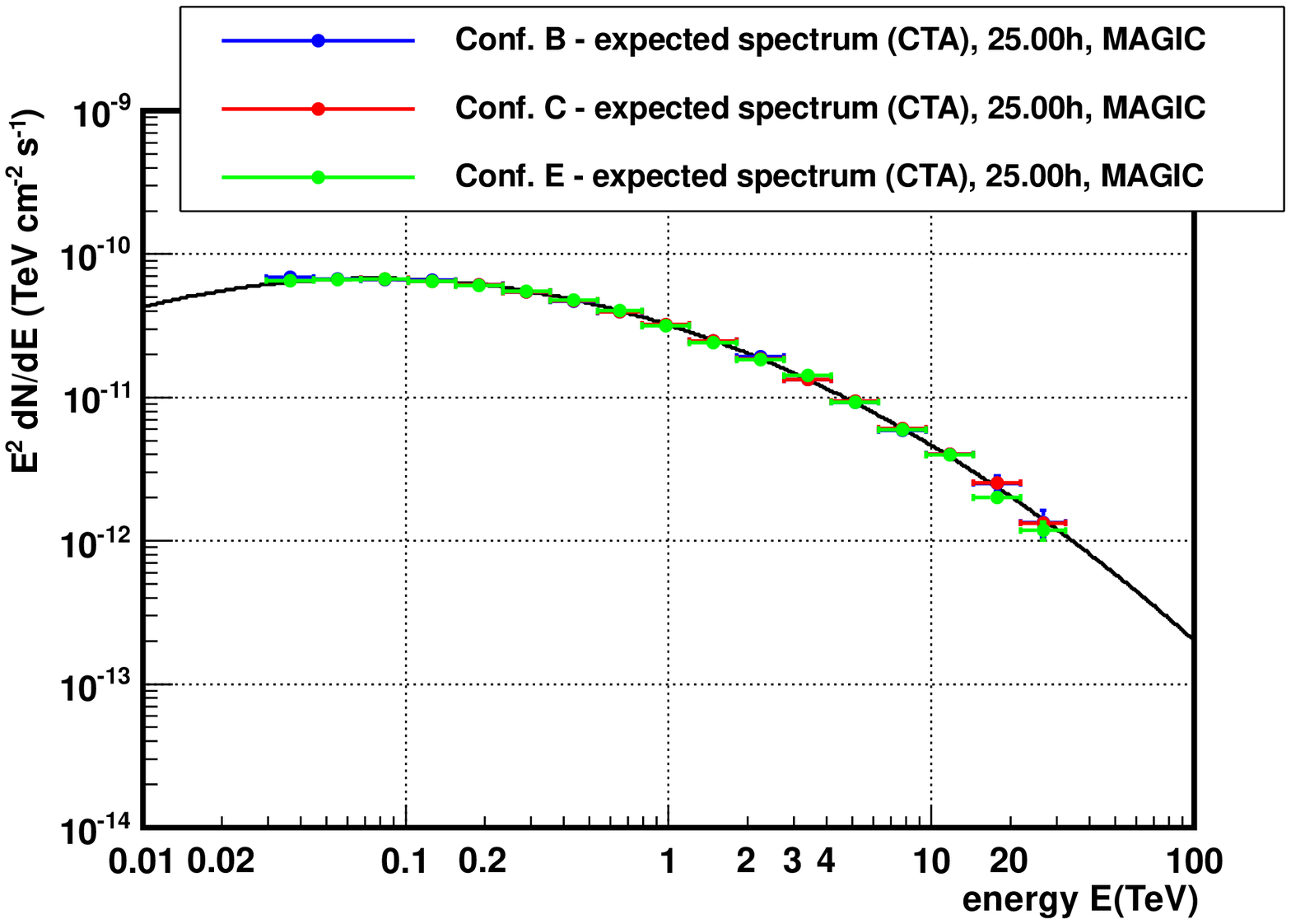} 
  \includegraphics[width=0.45\textwidth]{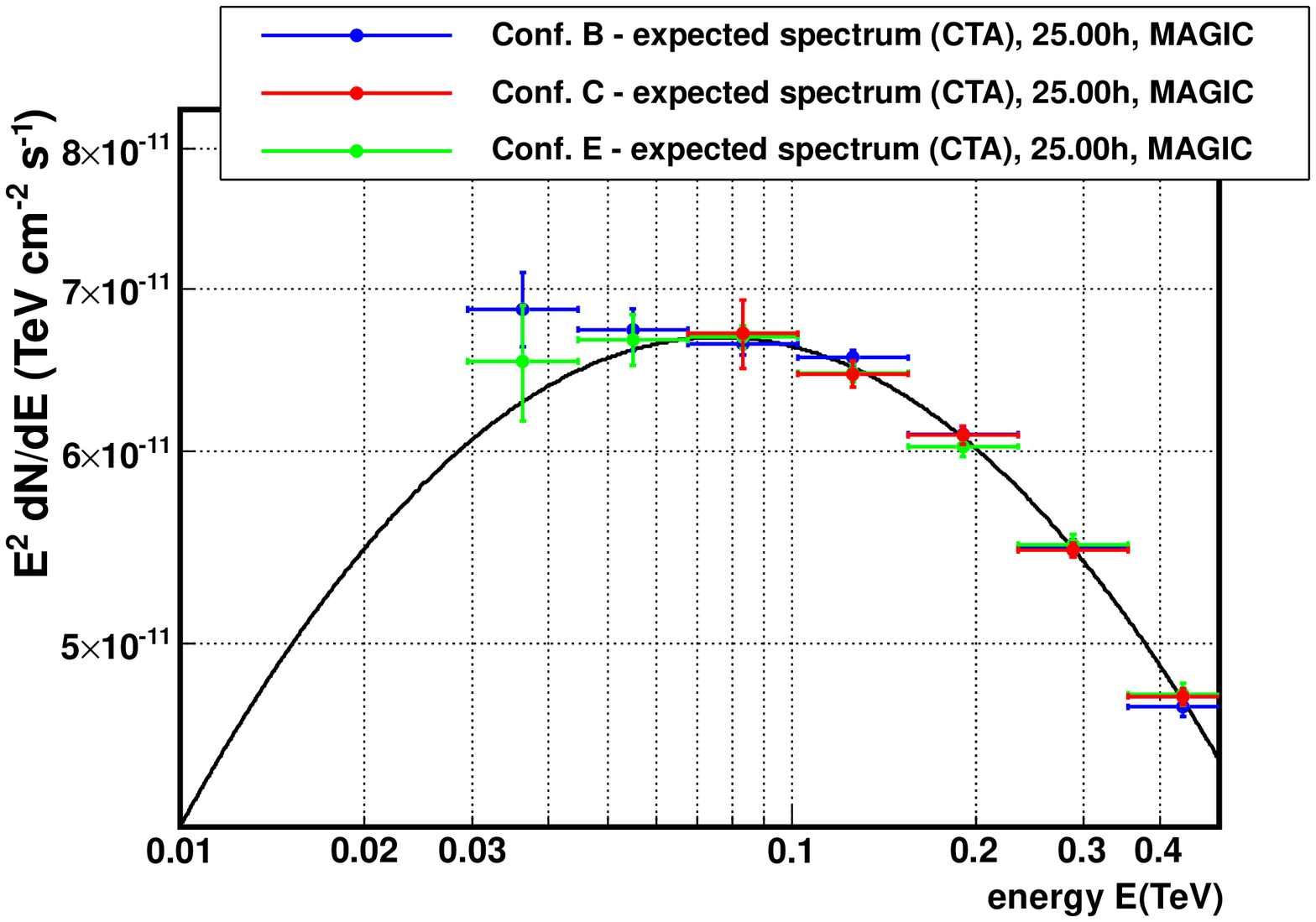} 
 \caption{On the left a) MAGIC-like reconstructed spectrum for the Crab Nebula using configurations B (blue), E (green) and C (red) in 25\,h. On the right b) zoom of the lowest energies for the MAGIC-like spectral shape}
 \label{fig2}
\end{figure}

Fig. \ref{fig2} a) shows the reconstructed spectrum for the three configurations using the MAGIC spectrum. A zoom of the lowest energies is shown on Fig. \ref{fig2} b). The original spectrum is shown in black.  The three configurations have similar performance although the energy threshold obtained for configuration B and E is lower (0.13\,TeV) than for C (0.32\,TeV), which is optimized to be more sensitive at high energies. Between the two first, the balanced configuration (E) seems to reconstruct more accurately the original spectrum at low energy (see Table \ref{table:1}). The simulations show that the spectrum at low energy where the IC peak is measured should be derived with high accuracy in a relatively short observation time, although, configurations B and E provide a lower energy threshold with respect configuration C, allowing a better determination of the IC peak position.

 \begin{figure}[t!]
 \centering
 \includegraphics[width=0.45\textwidth]{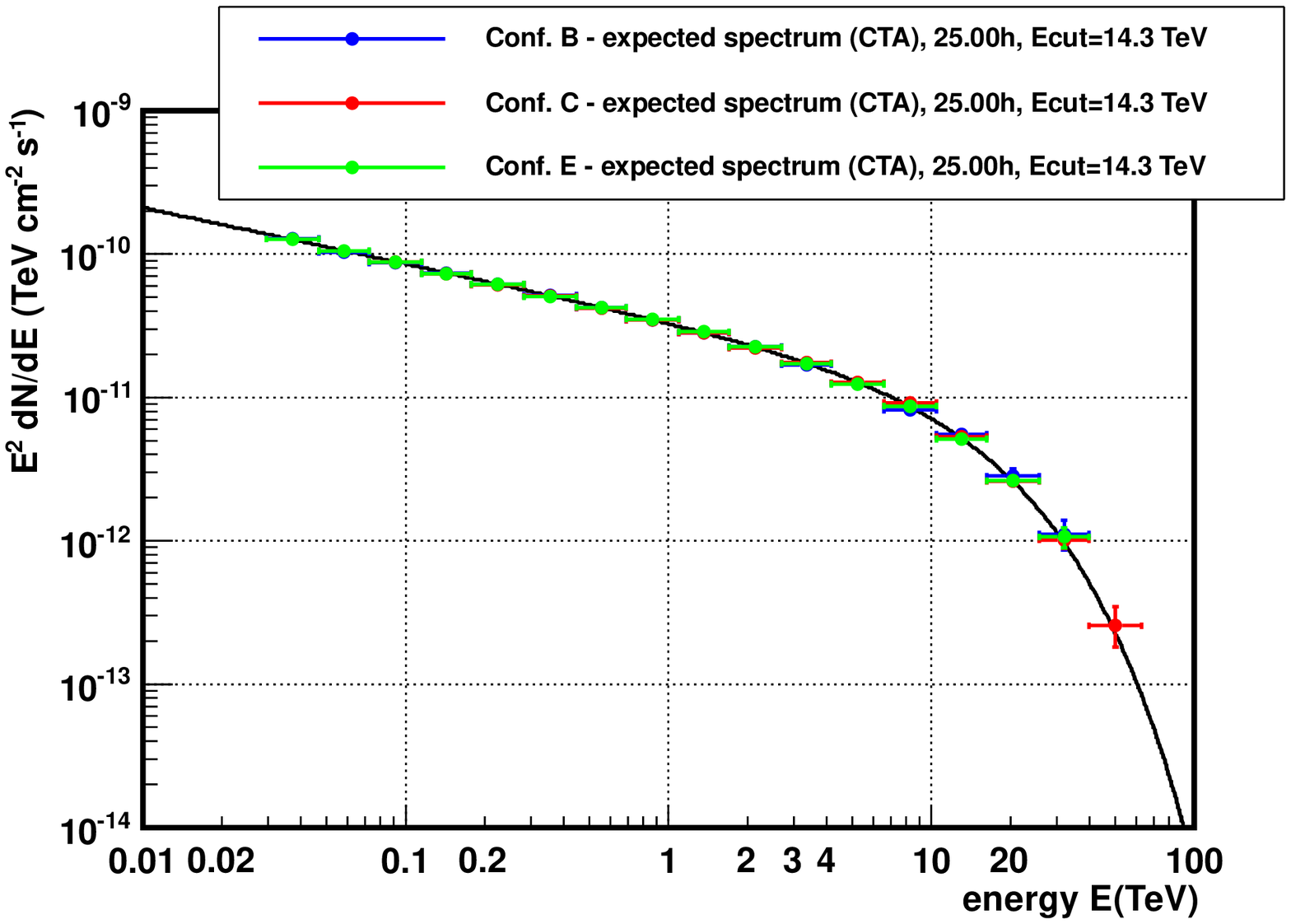} 
  \includegraphics[width=0.45\textwidth]{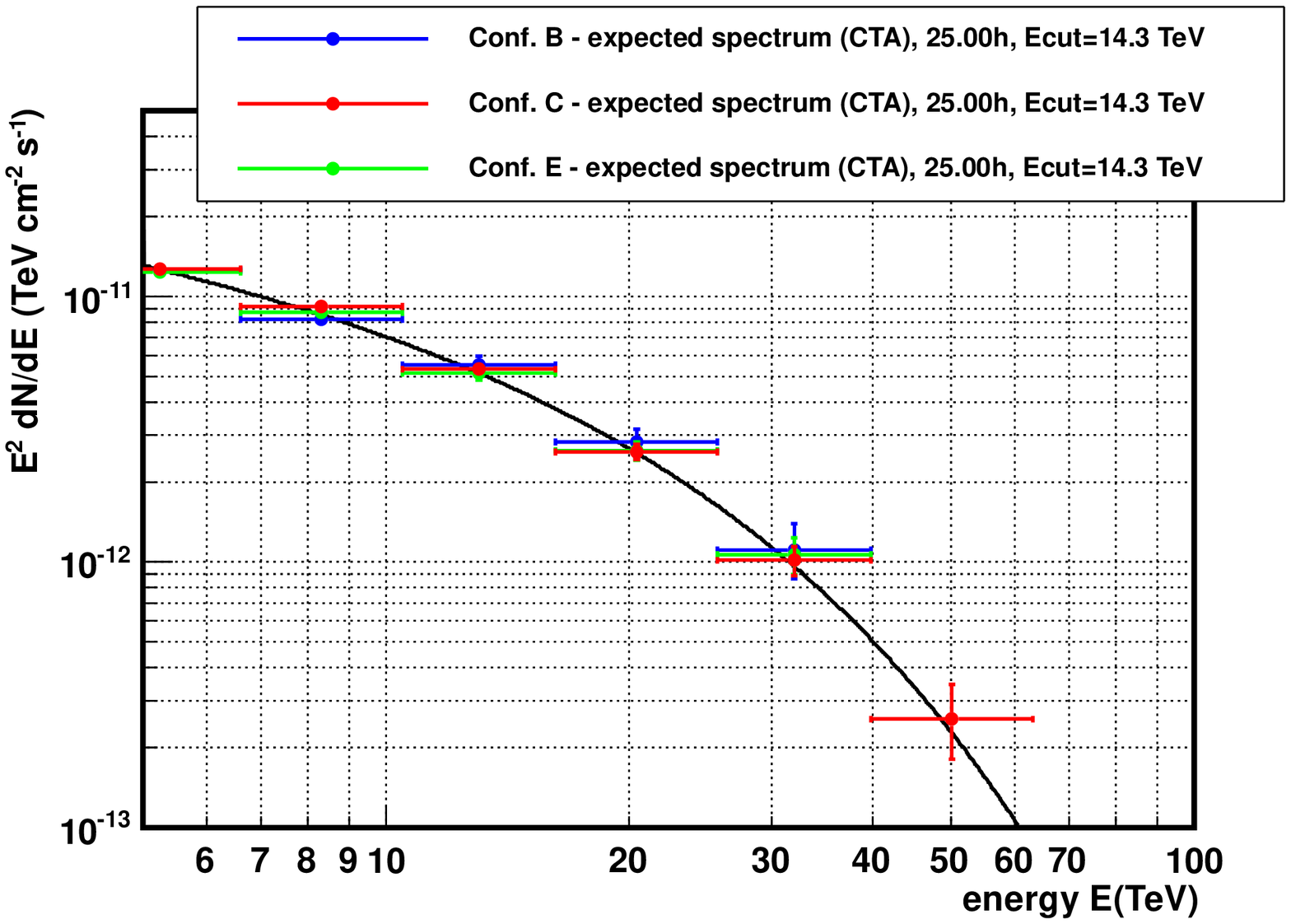} 
 \caption{On the left a) HESS-like reconstructed spectrum for the Crab Nebula using configurations B (blue), E (green) and C (red) on 25\,h. On the right b) zoom of the highest energies for the HESS-like spectral shape}
 \label{fig3}
\end{figure}

A similar analysis is shown in Fig. \ref{fig3}, using the HESS-like spectrum shape. A zoom of the cutoff region is displayed on the right b) (with fit values listed in Table \ref{table:1}). It shows clearly that similar to the previous case the three configuration reconstruct the high energy part of the spectrum correctly, with slight differences between them. Simulations with configuration C allows the extension of the spectrum up to 50\,TeV, which might be relevant with lower statistics. 

\begin{figure}[t!]
 \centering
 \includegraphics[width=\textwidth]{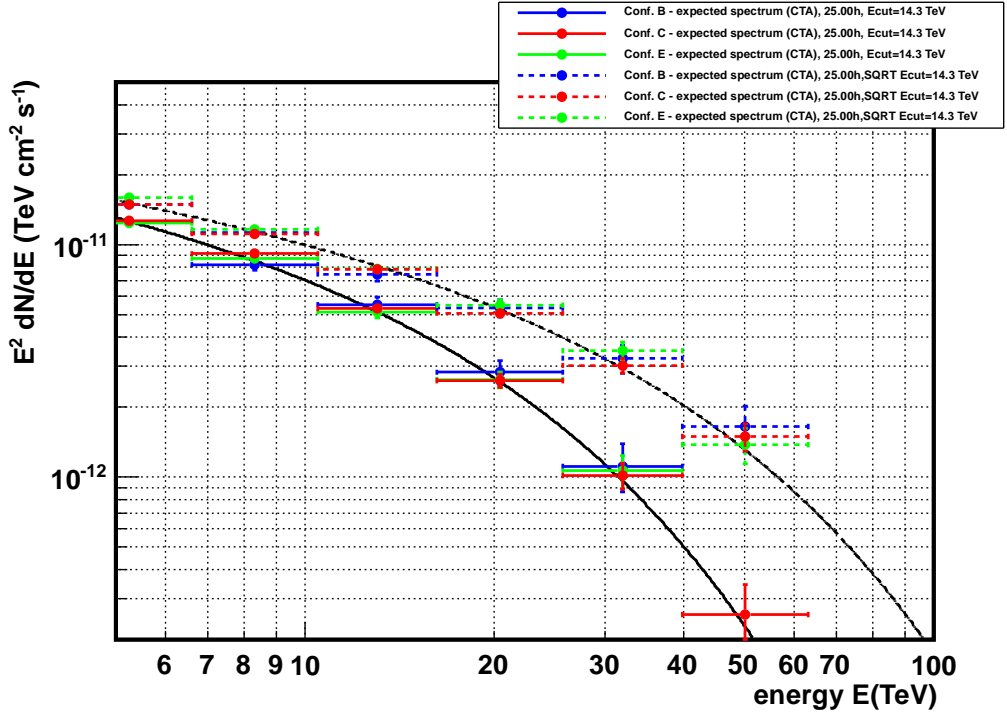} 
  \caption{HESS-like reconstructed spectrum for the two type of spectral cutoff considered (see text) for configurations B (blue), E (green) and C (red) for a simulated 25h observation . The solid line correspond to an exponential cutoff plus power-law spectral shape, whereas the dashed line represent a softer exponential cutoff.}
 \label{fig6}
\end{figure}

Concerning the variability observed at low energy (up to a few tens of GeV) it seems unlikely (given the minimum energy threshold obtained in 25h) to detect directly emission for eventual flares like the ones detected by AGILE and Fermi LAT. Nevertheless, if those flaring episodes are due to a change in the maximum energy reached of the underneath electron population or a change in its spectrum, it should be reflected on the IC emission at the highest energies. To evaluate the response of the different configuration to this type of effect, we simulate a softer cutoff, which translate in a harder spectral shape like dN/dE=I$_{\rm o}$(E/1 TeV)$^{-\Gamma}$[exp(-E/E$_{\rm cut}$)]$^{0.5}$ for an observation time of 25h. The reconstructed spectrum (see Fig. \ref{fig6}) is fit with the corresponding original spectra. 

\begin{center}
\begin{table}[t]
\caption{Fit values resulting from the simulation of a HESS-like and MAGIC-like Crab Nebula spectrum}
    \begin{tabular}{lccc}
   Configuration &  & I$_{\rm o}$ (10$^{-11}$ TeV$^{-1}$cm$^{-2}$s$^{-1}$) &    \\  	
    \hline	
HESS-like & E$_{\rm cut}$ & & $\Gamma$\\
  \hline	
    B  & 14.0$\pm$0.7& 3.52$\pm$0.03 & 2.385$\pm$0.004 \\
    C  & 14.7$\pm$0.4 & 3.48$\pm$0.02 & 2.388$\pm$0.004 \\
    E  &  14.3$\pm$0.6& 3.50$\pm$0.03 & 2.387$\pm$0.004 \\
    \hline
    MAGIC-like & b & & a \\
      \hline	
    B  & 0.249$\pm$0.005  & 0.600$\pm$0.002 & 2.312$\pm$0.004 \\
    C  & 0.255$\pm$0.005 & 0.601$\pm$0.002 & 2.307$\pm$0.005 \\
    E  & 0.260$\pm$0.005 & 0.601$\pm$0.003 & 2.308$\pm$0.004 \\
    \end{tabular}    	
\label{table:1}
\end{table}
\end{center}

\begin{figure}[t!]
 \centering
 \includegraphics[width=0.45\textwidth]{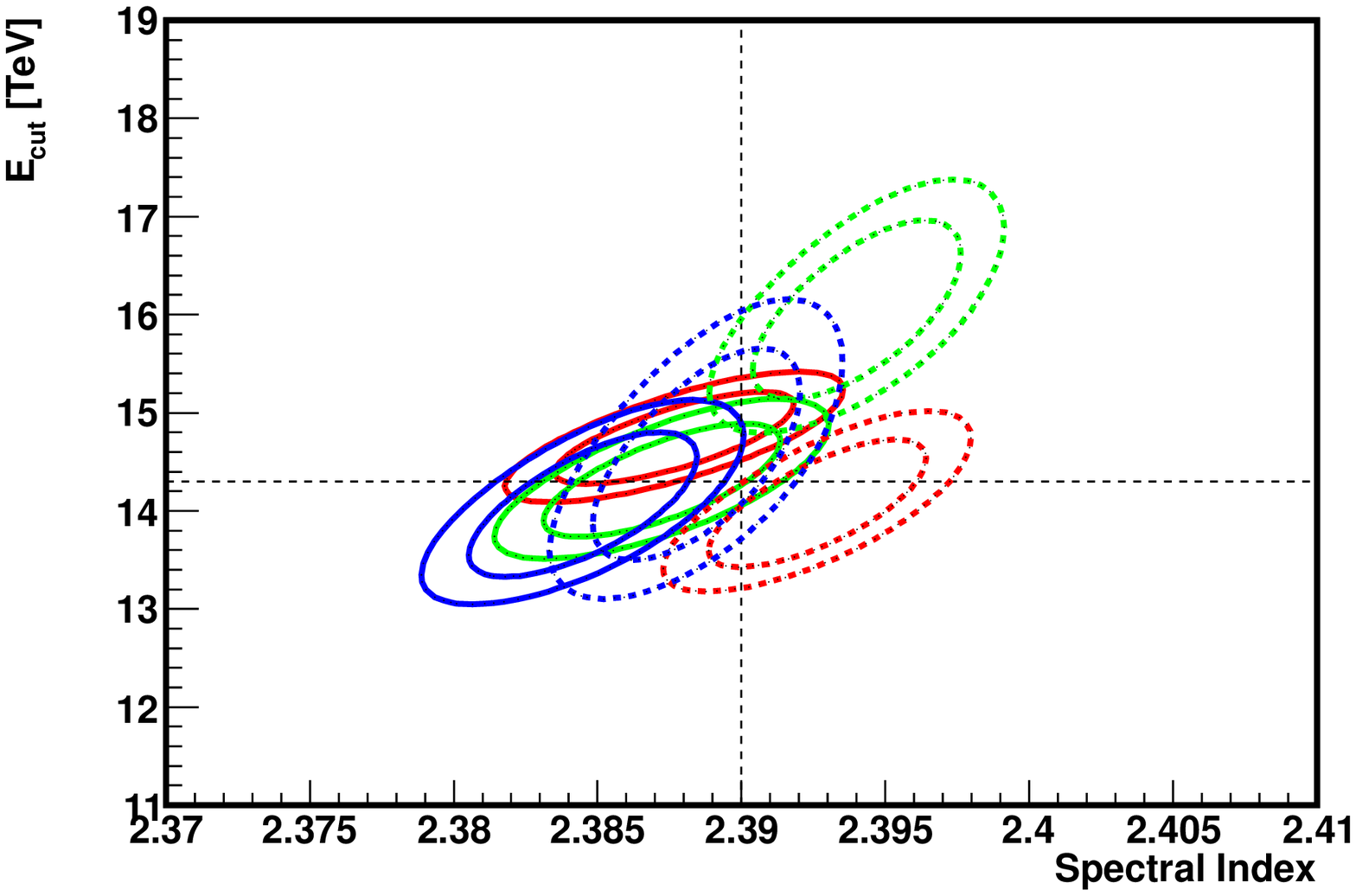} 
  \includegraphics[width=0.45\textwidth]{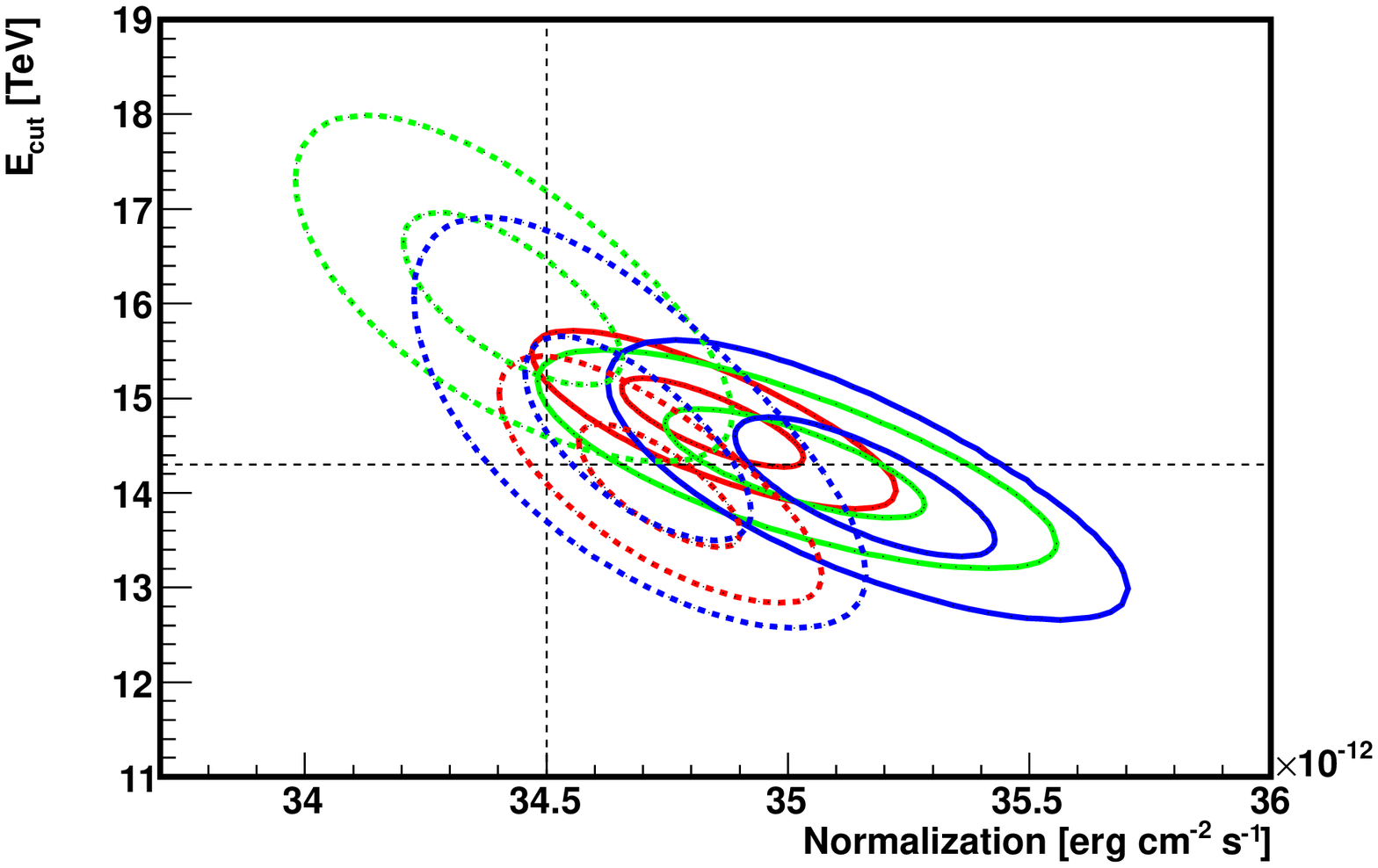} 
 \caption{1 and 2$\sigma$ contours for the fit parameters to the measured Crab Nebula using configuration B (blue), E (green) and C (red) for the HESS-like spectrum using an exponential energy cutoff (solid lines) and a softer one (dashed lines).}
 \label{fig7}
\end{figure}

Fig. \ref{fig7} shows the 1 and 2$\sigma$ contours for the best-fit parameters. Configuration C and B reconstruct the spectral parameters correctly for the two type of functions used while configuration E seems to have a worse response when deriving the energy cut. Nevertheless, the three configuration lead to compatible results and a hardening of the Crab Nebula spectrum should be detectable with any of the three proposed configuration in less than 25h.

Finally we check the capability of CTA to detect changes in the Crab Nebula size due, for instance, to cooling effects. In that scenario, a enlargement of the size (IC produced by low energy electrons around the compact PWN) could be expected at very low energy when comparing with the highest energies. We used again the B, E and C configurations and considered the case \emph{cooling} in {\it CTAmacrosv6.C} with soft ($\Gamma$=2.7) Gaussian ($\sigma$=0.1) point-like source and a second slightly smaller ($\sigma$=0.07) Gaussian-shaped source with harder spectral index ($\Gamma$=2.0) for an observation time of 25~h. The resulting photon image is then fitted with a Gaussian function convolved with the configuration point spread function (PSF) and the fit sigma is recovered for two energy bands, a low energy range between 0.05 and 0.1 GeV and a high energy range between 1 and 10 TeV. The results are listed in Table \ref{table:2}.
At high energies, the Gaussian is well reconstructed for all configurations and Crab Nebula flux level, configuration C being the one that reports smaller errors in the fitted sigma. On the contrary, at low energies, the flux has to be increased at least up to 10 Crab unit (c. u.) (for 25 h observation time) in case of configuration C and E, and up to 5 c. u. for configuration B (as expected since B should be optimized for low energy events).

\begin{center}
 \begin{table}[t]
\caption{Fit values resulting from cooling effects simulation for a Crab-like source, for 25~h observation time. At low energies (LE) the reconstructed Gaussian sigma should be 0.1$^{\rm o}$ whereas at high energies (HE) a shrink should be observed ($\sigma$=0.07$^{\rm o}$).}
\resizebox{13cm}{!} {
    \begin{tabular}{ccccccc}
   Conf. & 1 c. u. & 5 c. u. & 10 c. u. & 50 c. u. & 100 c. u. & PSF\\  	
\hline
	HE & & & &  &\\
    \hline	
    B  & 0.07$\pm$0.04 & 0.07$\pm$0.01  &  0.073$\pm$0.007  & 0.072$\pm$0.003 & 0.072$\pm$0.002  & 0.046 \\
    C  & 0.07$\pm$0.02 & 0.072$\pm$0.006  &  0.072$\pm$0.004  & 0.072$\pm$0.002 & 0.072$\pm$0.002  &  0.058 \\
    E  & 0.07$\pm$0.03 & 0.073$\pm$0.009  &  0.072$\pm$0.005  & 0.072$\pm$0.002 & 0.072$\pm$0.002  &  0.058 \\
    \hline
    	LE & & & &  &\\
    \hline	
    B  & 0 & 0.14$\pm$0.05 & 0.10$\pm$0.03 & 0.11$\pm$0.01 & 0.104$\pm$0.007 & 0.163\\
    C  & 0 & 0 & 0.1$\pm$0.3 & 0.1$\pm$0.1 & 0.10$\pm$0.08 & 0.304 \\
    E  & 0 & 0 & 0.2$\pm$0.2 & 0.14$\pm$0.08 & 0.13$\pm$0.05 & 0.304 \\
    \end{tabular}    	}
\label{table:2}
\end{table}
\end{center}

\subsubsection{PWNe Morphology}

\begin{figure}[t!]
 \centering
 \includegraphics[width=0.49\textwidth]{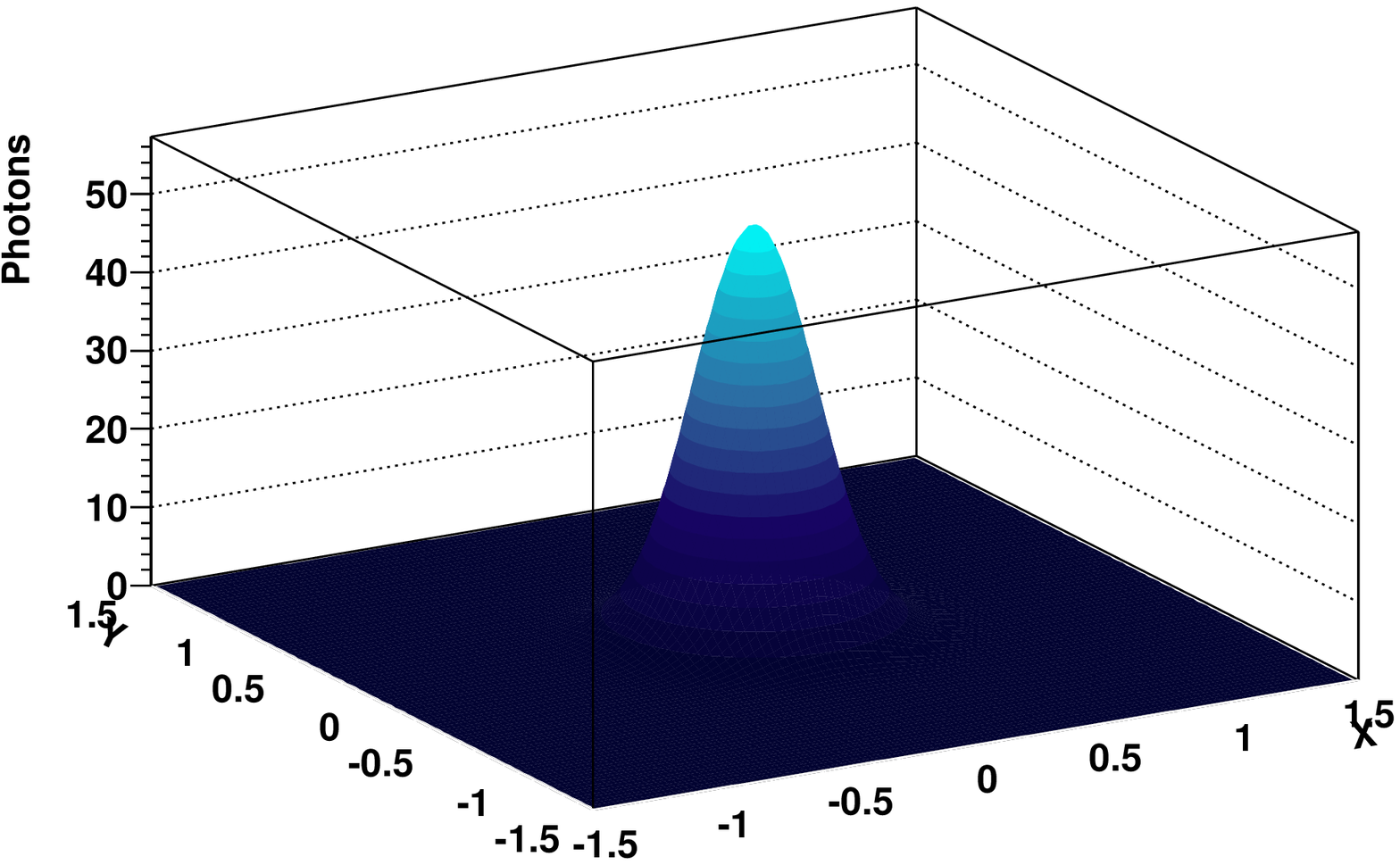} 
  \includegraphics[width=0.49\textwidth]{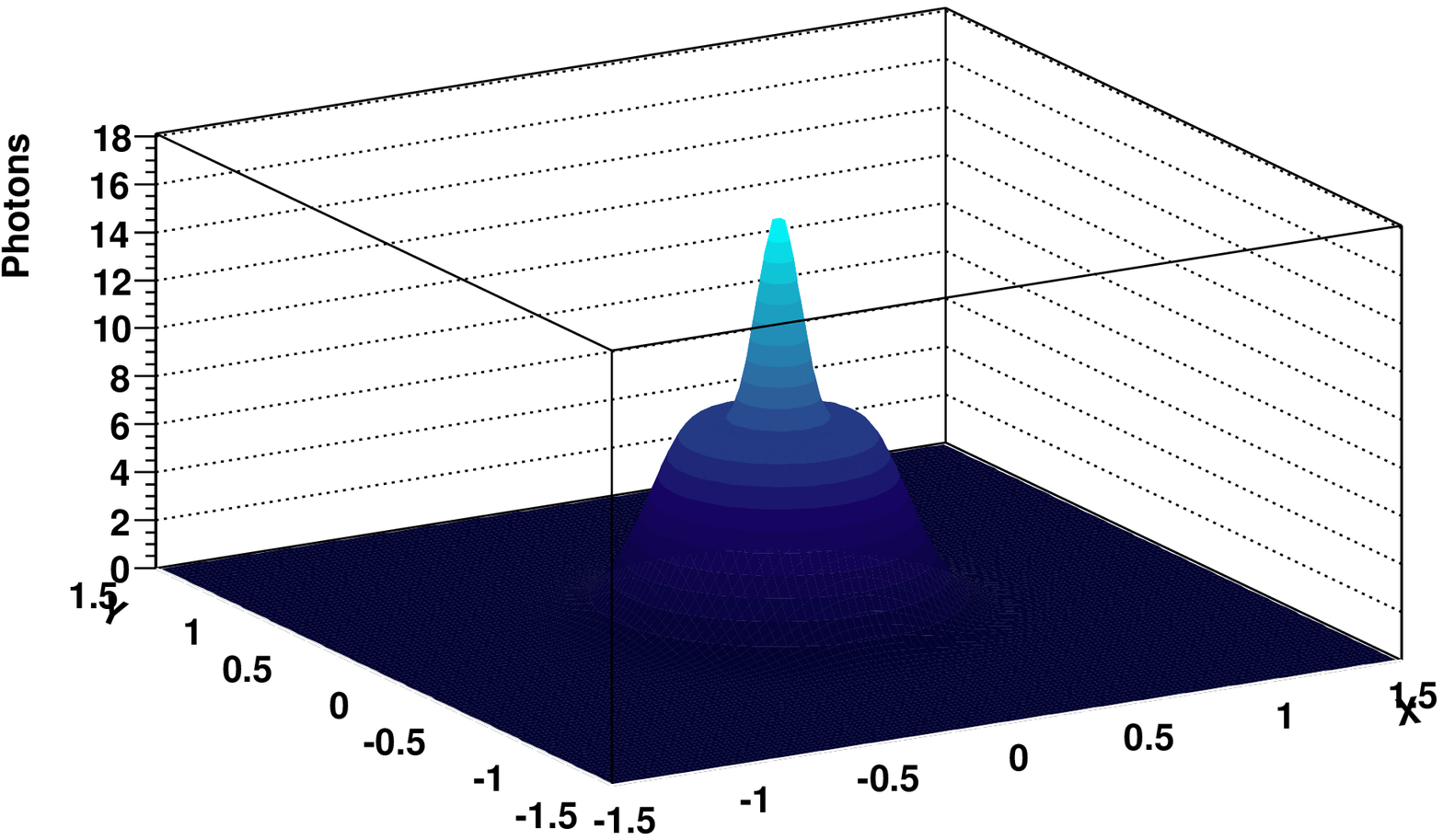} 
 \caption{Gaussian-shape and composite SNR-shape simulations for a source with a Crab-like spectrum as detected with CTA Configuration E after 50 hours}
 \label{fig8}
\end{figure}

\begin{figure}[t!]
 \centering
 \includegraphics[width=\textwidth]{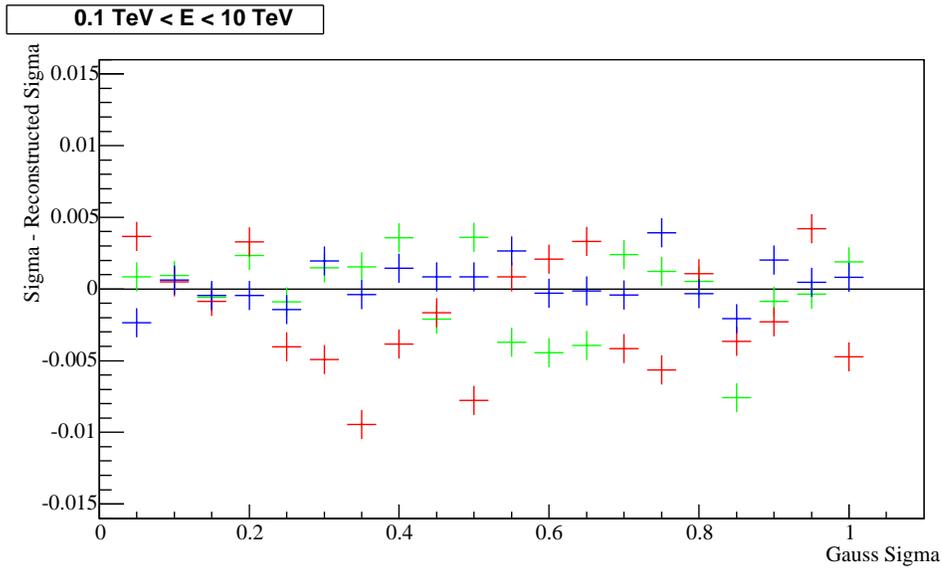}
 \caption{Gaussian extension reconstructed for different configurations (B in red, C in blue and E in green) for a 1 c. u. source in 25h}
\label{fig9}
\end{figure}

To assess the capability of CTA to resolve the morphology of PWNe, different VHE $\gamma$-ray shapes for given spectral inputs were simulated. The $\gamma$-ray flux is then convolved with the CTA response and the VHE $\gamma$-ray excess is recovered after subtracting the cosmic-ray background. Fig. \ref{fig8} shows two particular cases. On the left, the excess as it will be observed with CTA for configuration E for a 2-dimension Gaussian-shape source is displayed whereas the right panel shows a second source composed by a Gaussian point-like source surrounded by a shell-like excess (composite SNR). The reconstructed sigma of different Gaussian-shape excesses do not show large uncertainties for the different configurations (or different energy bands). Fig. \ref{fig9} shows one example (1 c. u., 25\,h) for the three configurations (in green, blue and red for configurations E, C and B respectively) in the 1 to 10\,TeV energy band.

To evaluate the response of the different configuration for very large sources, we simulated a Gaussian source with $\sigma$=1.2$^{\rm o}$, similar to the extension observed by H.E.S.S. in Vela X \cite{velax}. The results show a precise reconstructed Gaussian $\sigma$ with errors of the order of 0.001 for all configurations. 

\begin{figure}[t!]
 \centering
 \includegraphics[width=0.48\textwidth]{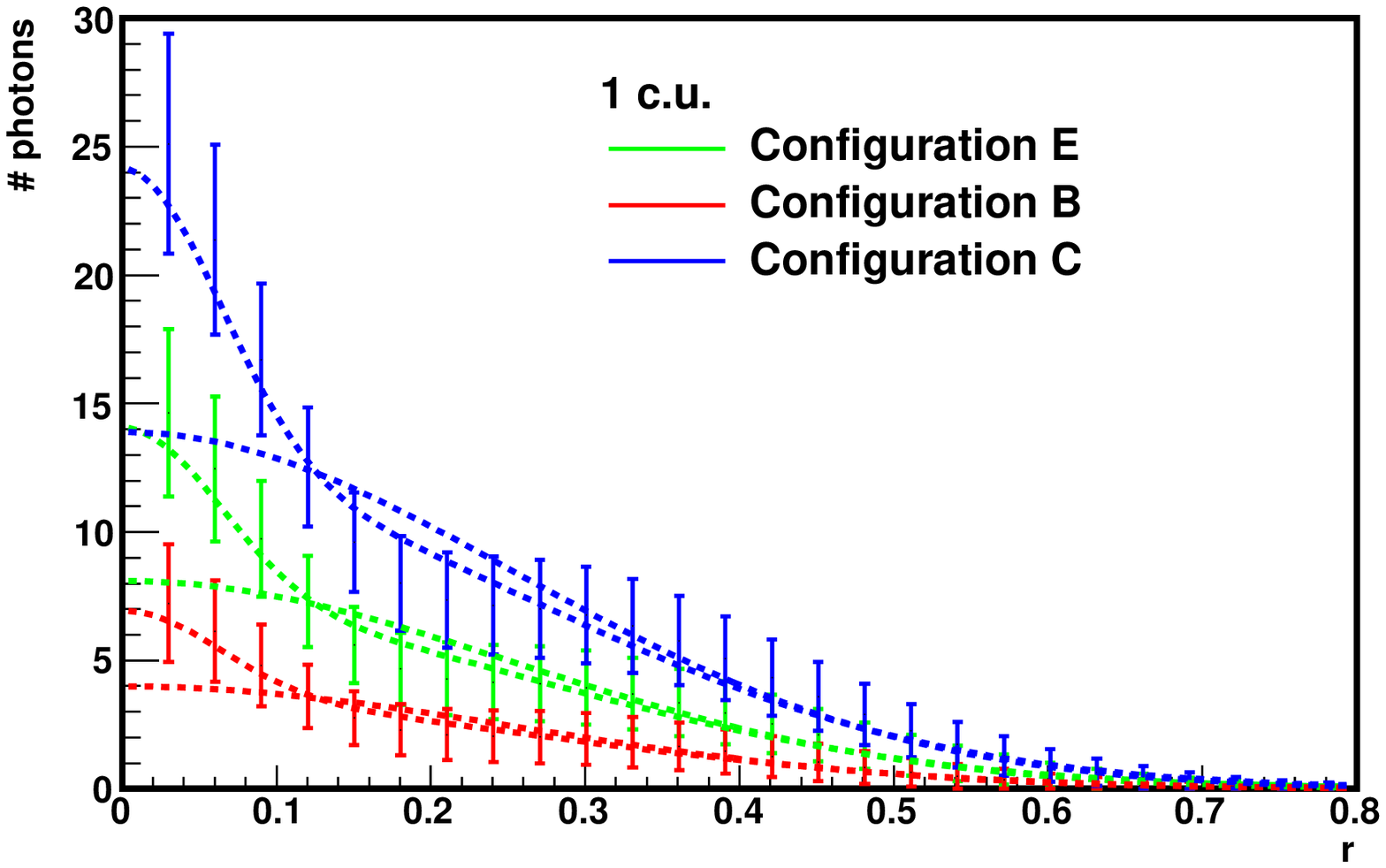}
  \includegraphics[width=0.48\textwidth]{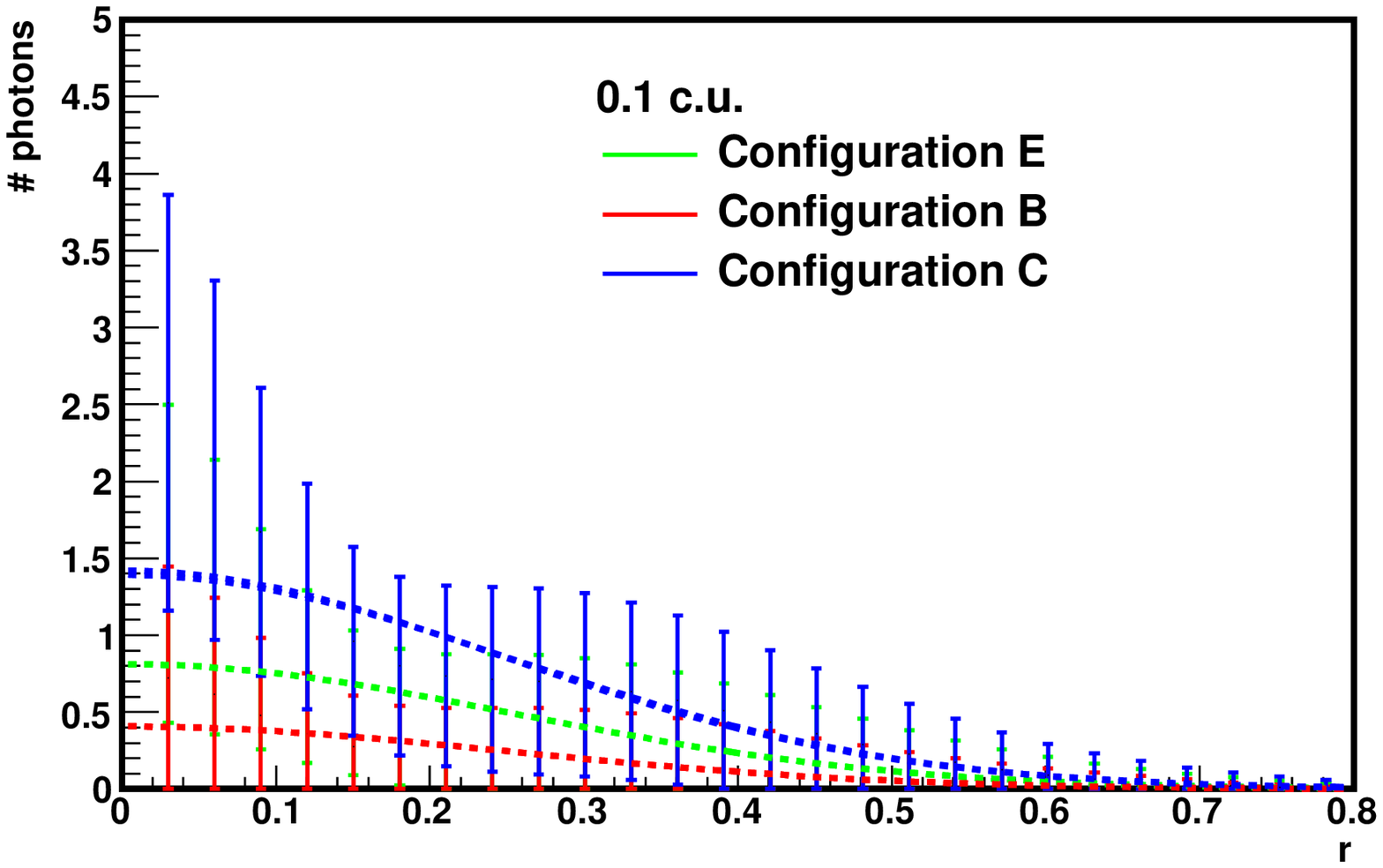}
 \caption{Radial profiles derived for a composite SNR simulation (B in red, C in blue and E in green). The dashed lines show the fits to a single and a double Gaussian function for each 50h simulation.}
\label{fig10}
\end{figure}

Finally, the radial profile of a simulated composite SNR is derived for the three configurations and two integral fluxes, 1 and 0.1 c. u. in 50~h. For a composite SNR at 1 c. u., a fit to double Gaussian as opposite to a single one fits better the data for the three configurations (in the whole CTA energy range from 0.05 to 100 TeV) (Fig. \ref{fig10} left) whereas for 0.1 c. u. in 50 h, the components are only clearly distinguished for configuration C (in blue in Fig. \ref{fig10} right). 

In conclusion, no large differences are observed, being the configuration C the most appropriated for composite SNRs.

\subsubsection{Cooling effects}
CTA observations were simulated in order to study energy-dependent morphologies in extended sources. 
The case of HESS\,J1825--137 is used as an example of morphology shaped by synchrotron cooling. The reference paper is \cite{1825}, from which we take the surface brightness and spectral index distribution. We also assume that those distributions can be extended with the same behavior out to larger radii than those probed by the H.E.S.S. observation.

Configuration B and E were used, for which offset-dependent Monte Carlo simulations are available. As the majority of the investigated spectra are rather steep, we anyway expect these two to be the most important configurations (configuration D/C are optimized for high energies).

In order to simulate the spectral evolution and maps, we simulate similar wedges to those used in \cite{1825}. For each wedge, the photon spectrum is a power-law ($dN/dE=N_0 (E/E_0)^{-\Gamma}$) with spectral index $\Gamma$. 

\begin{figure}[!htbp]
\begin{center}  
\includegraphics[width=0.49\linewidth]{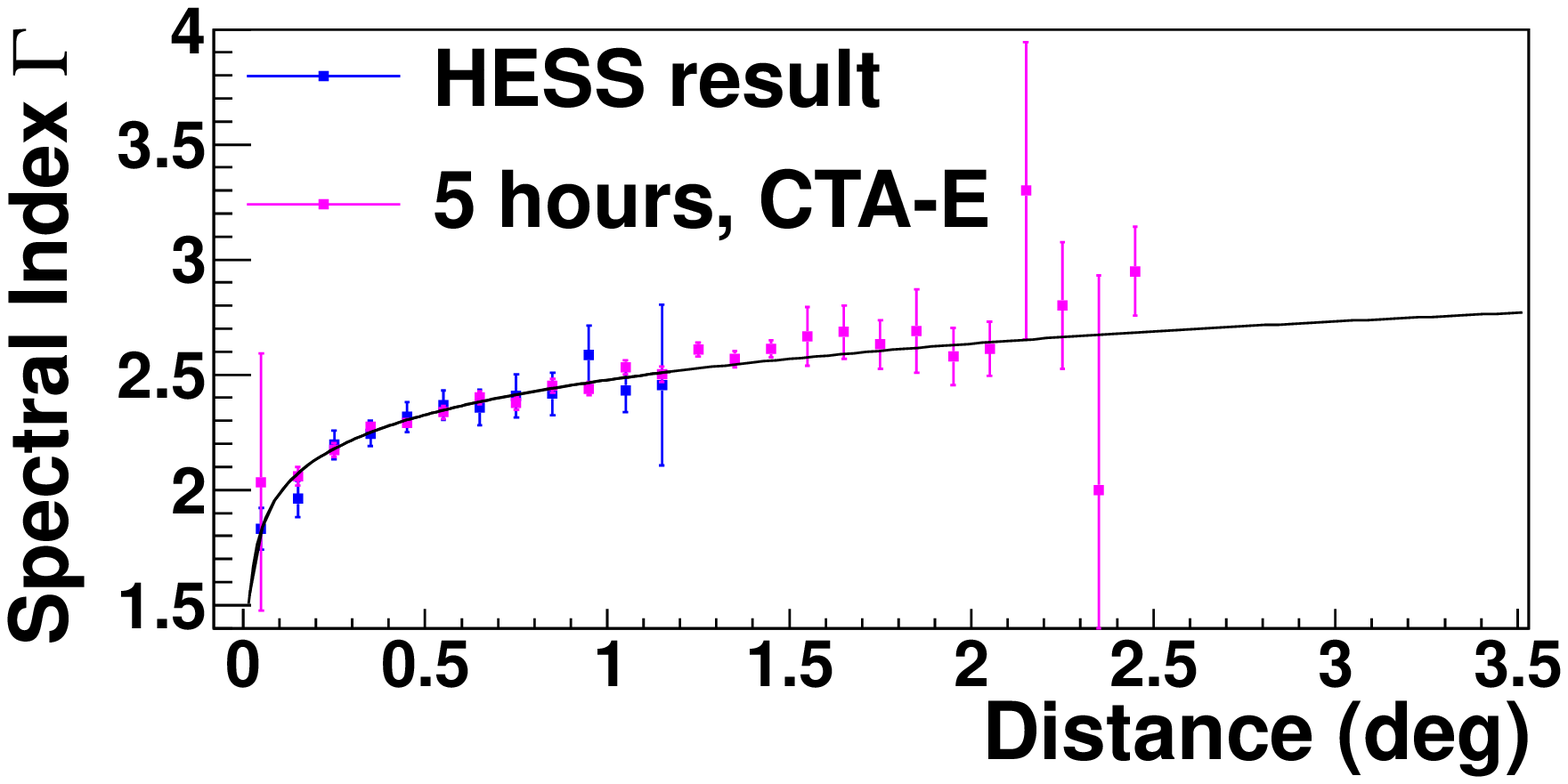}
\includegraphics[width=0.49\linewidth]{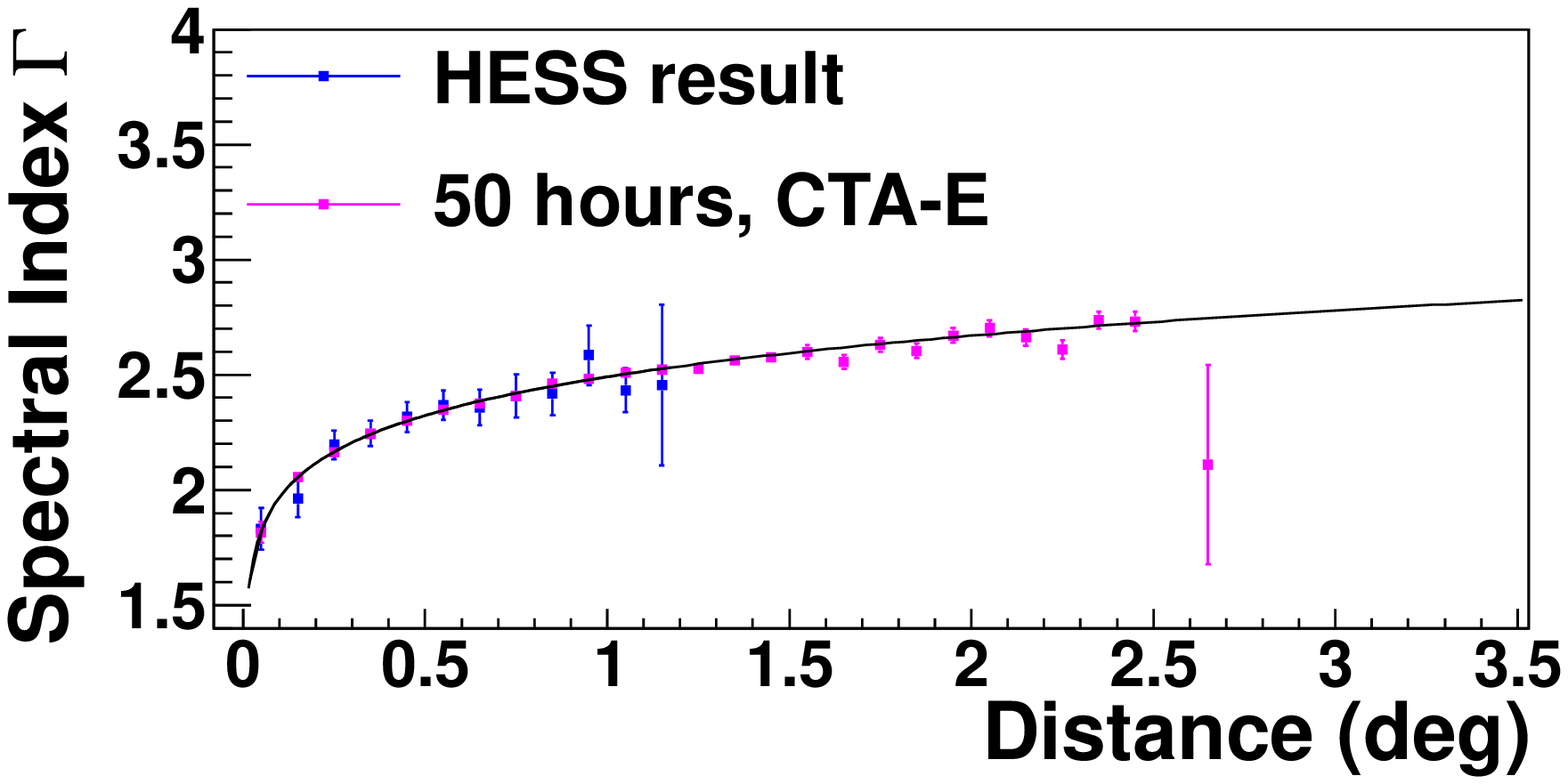}
  \caption{Simulated spectral index as a function of distance from the center of the object in magenta point. The assumed intrinsic distribution is derived from the H.E.S.S. data (blue) and it is shown as a black solid line. On the left a) 5h of observations with configuration E. On the right b) 50h of observations with configuration E.}
 \label{fig11}
\end{center}
\end{figure}

\begin{figure}[!htbp]
\begin{center}  
\includegraphics[width=\linewidth]{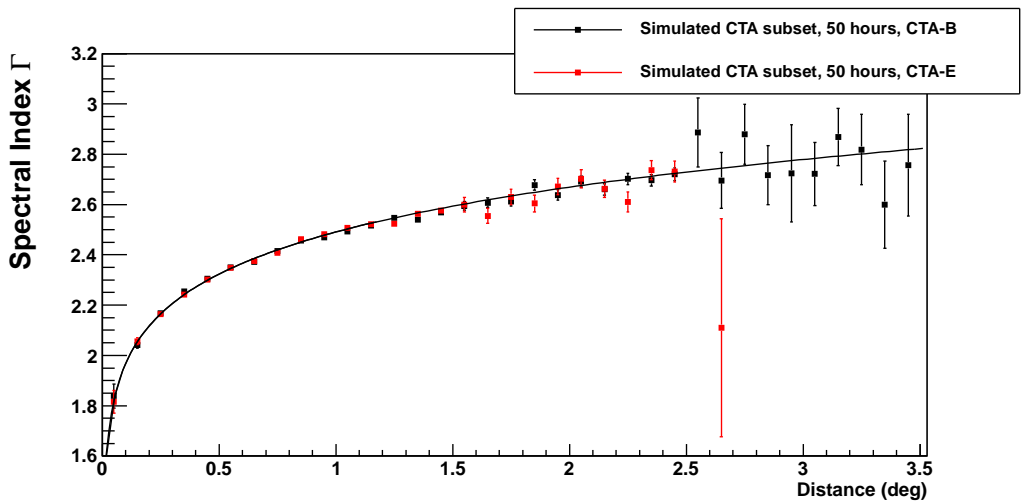}
  \caption{Reconstructed spectral index for a 50h observation signal using configuration B vs configuration E. Below 2.5$^{\rm o}$ the two distributions are compatible at the 63\% level. }
 \label{fig12}
\end{center}
\end{figure}

Examples of the simulated index distribution are shown in Fig. \ref{fig11}. It can be seen that with the assumed index and surface brightness distribution, E will be able to probe out to a 2.5$^{\rm o}$ distance with only 5h of observation time. Within 50h, the error on the index determination is much reduced, but we will not be able to probe out to a much larger extension. Configuration B will probe out to larger extension and with a better error estimation on the photon index (see Fig. \ref{fig12}). This is due to the better sensitivity at lower energies, necessary for the rather steep spectra of the outer shells.

\begin{figure}[!htbp]
\begin{center}  
\includegraphics[width=\linewidth]{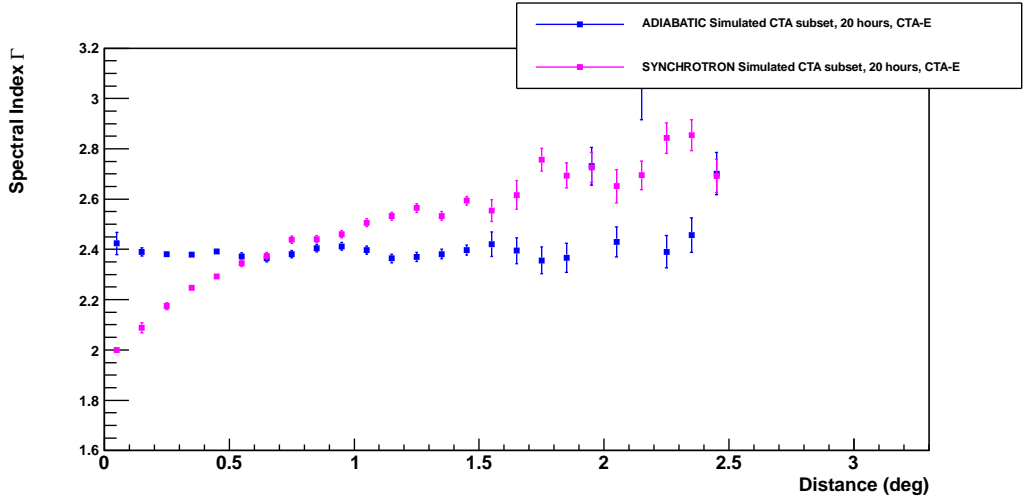}
  \caption{Reconstructed spectral index for 20h observation simulated signal using configuration E. The two simulated cases, synchrotron or adiabatic cooling show a clear distinguishable spectral index distribution}
 \label{fig13}
\end{center}
\end{figure}

We can also simulate the capability of CTA to disentangle different spectral indices for the same surface brightness distribution. In 20h observation time it is possible to distinguish among scenarios of synchrotron cooling or adiabatic cooling (constant spectral index). This is shown in Fig. \ref{fig13} where x refers to the angular distance to pulsar.


\begin{figure}[!htbp]
\begin{center}  
\includegraphics[width=0.49\linewidth]{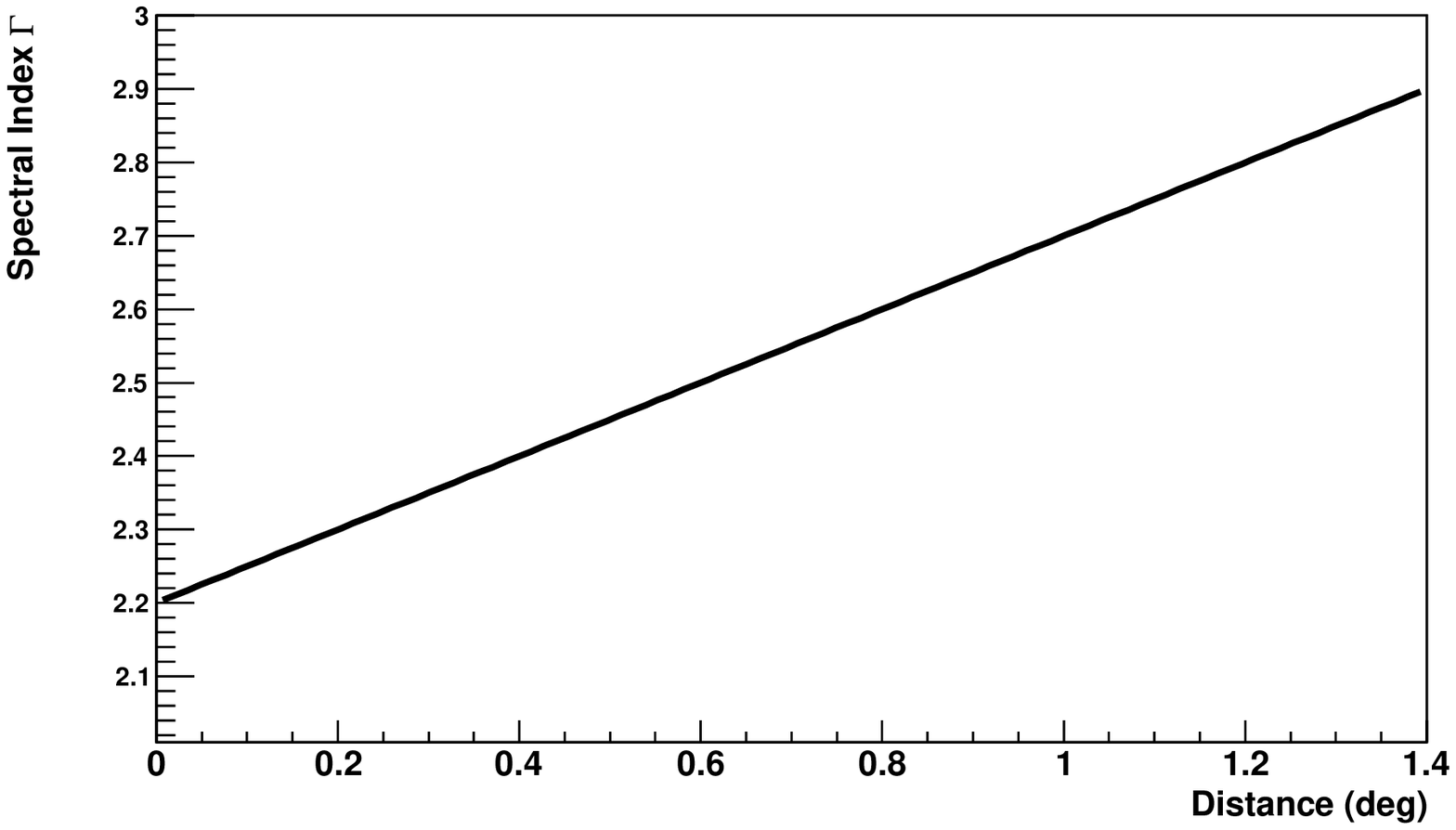}
\includegraphics[width=0.49\linewidth]{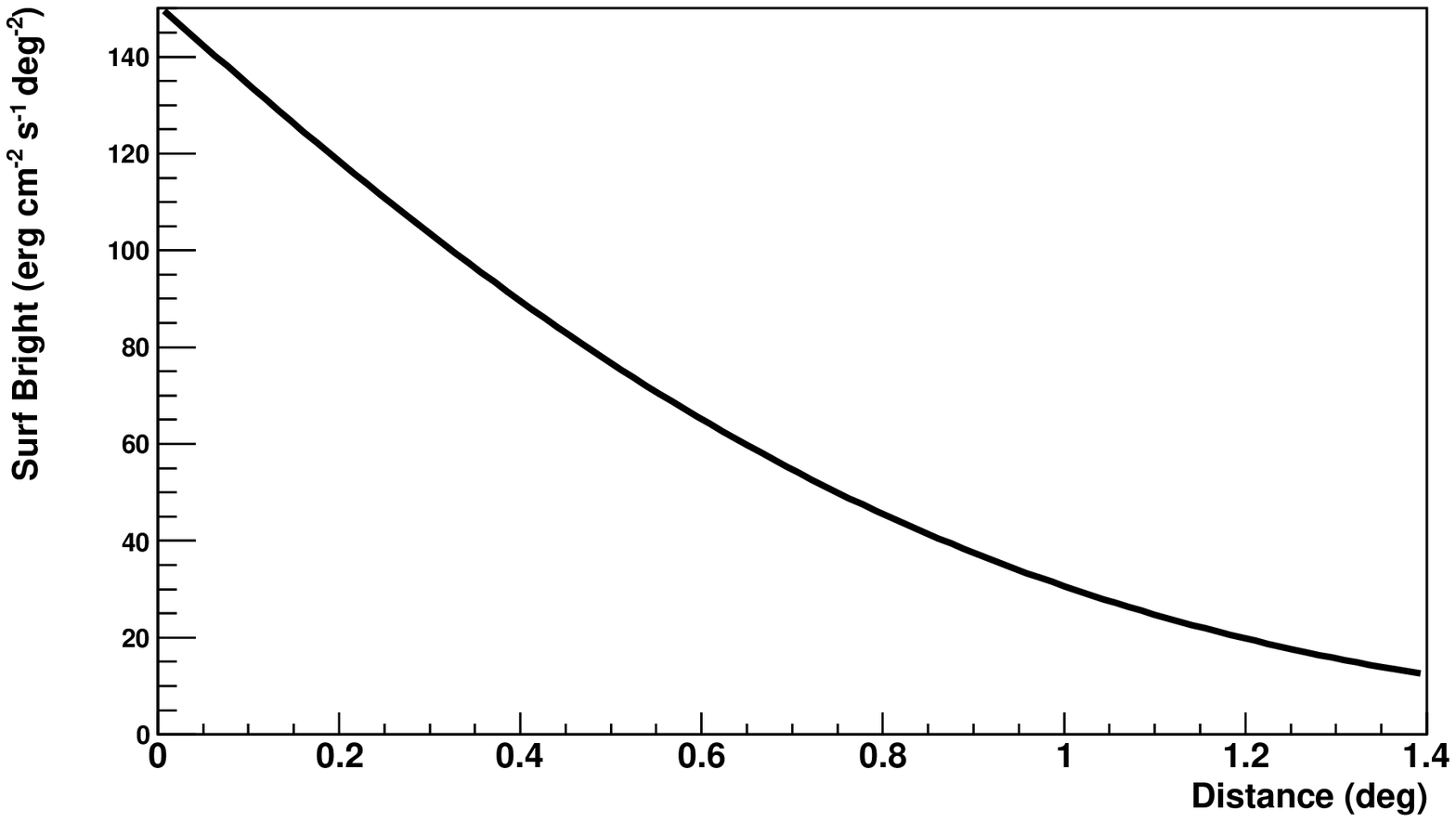}
  \caption{Assumed distributions of index (left) and surface brightness (right) for the general case of cooling.}
 \label{fig14}
\end{center}
\end{figure}

\begin{figure}[!htbp]
\begin{center}  
\includegraphics[width=0.49\linewidth]{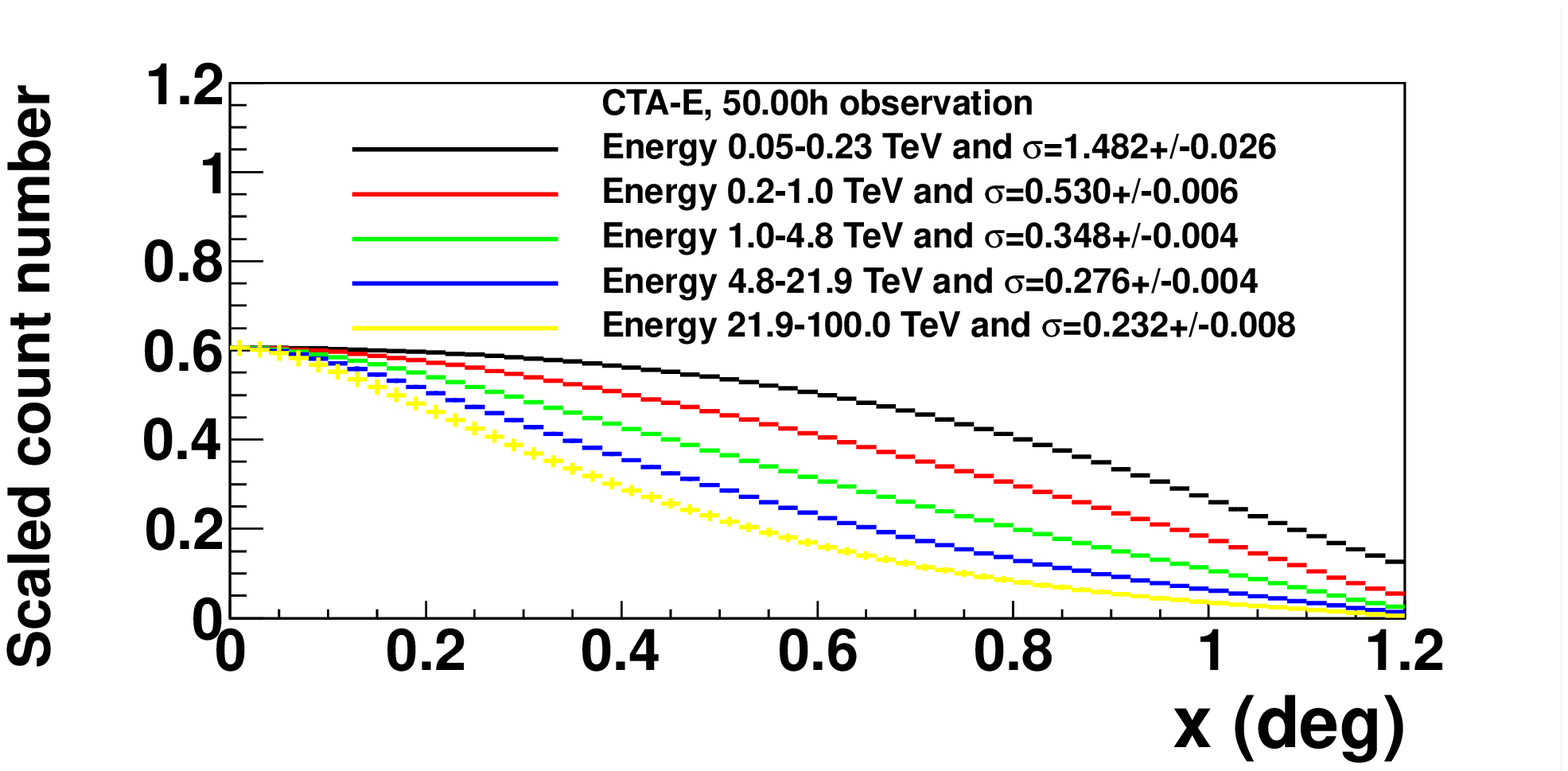}
\includegraphics[width=0.49\linewidth]{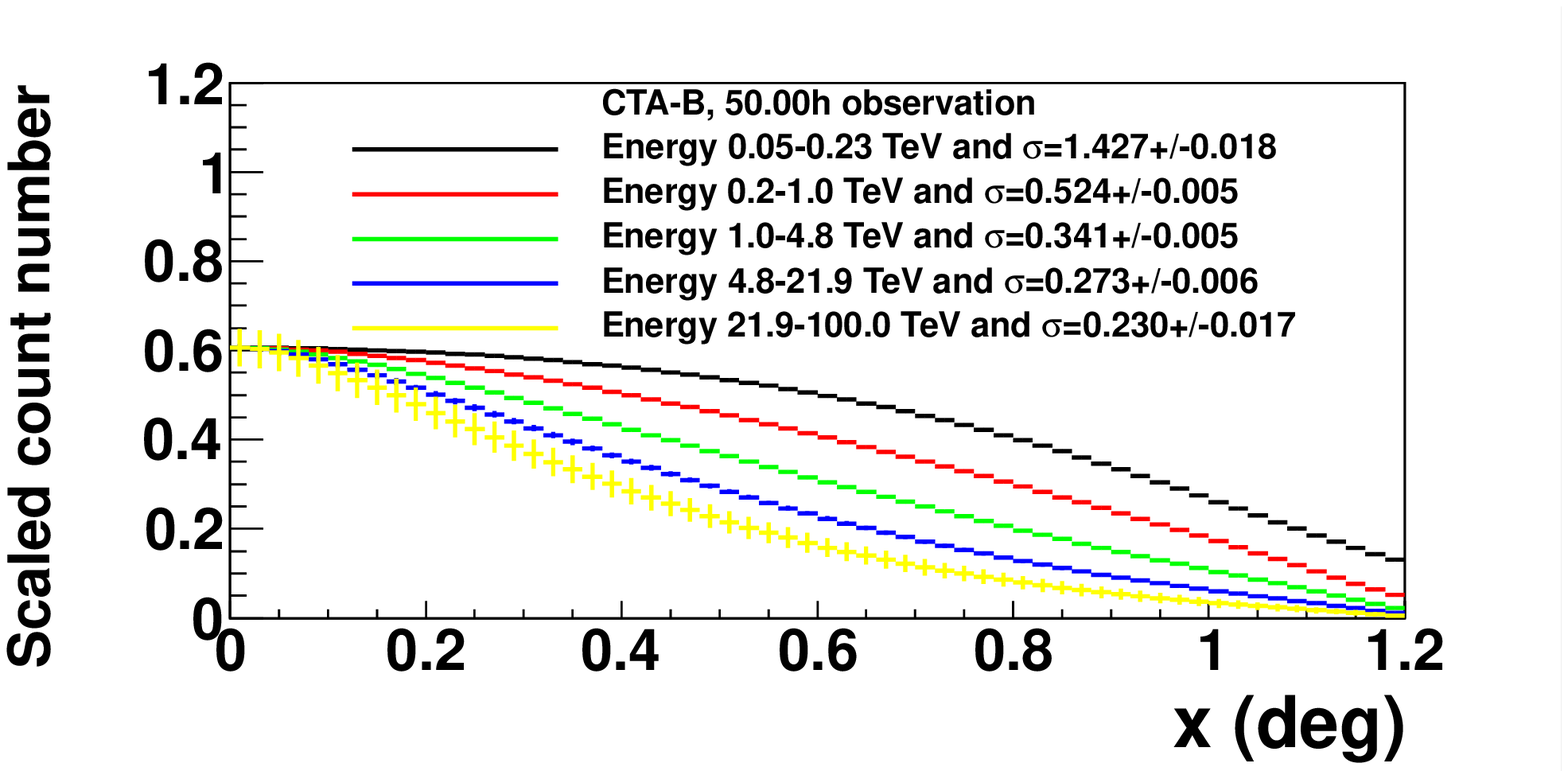}
\includegraphics[width=0.49\linewidth]{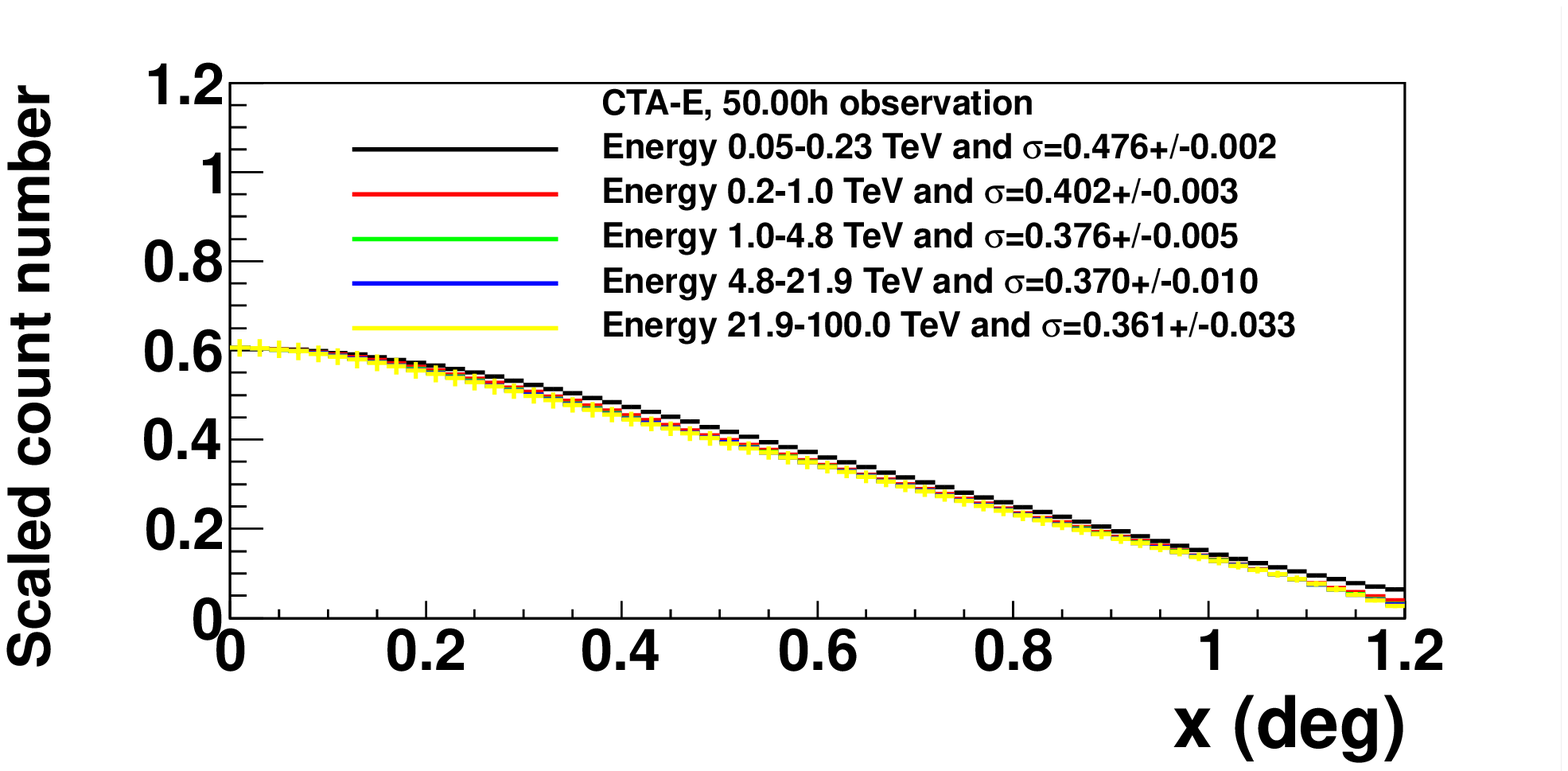}
\includegraphics[width=0.49\linewidth]{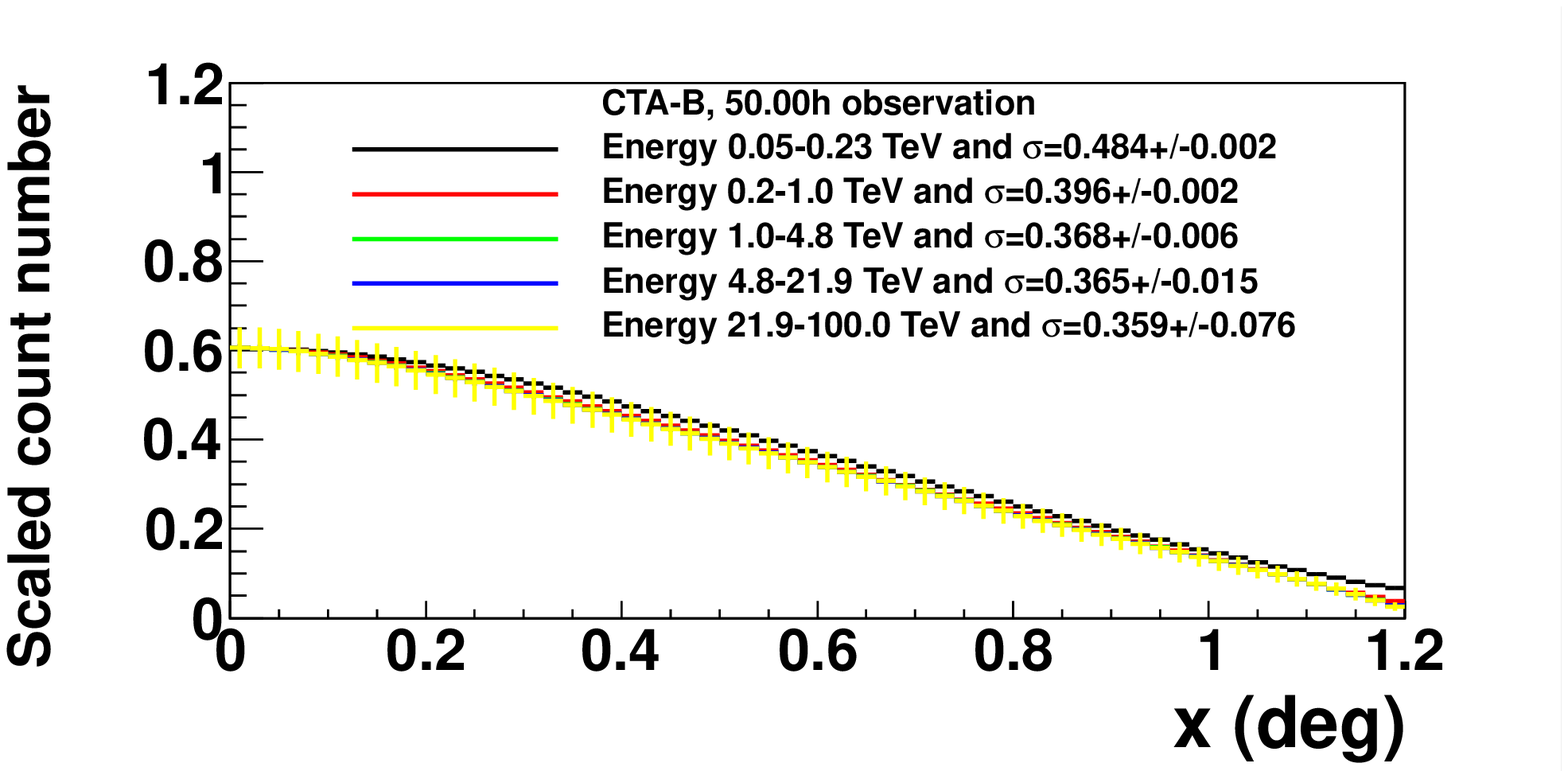}
\caption{Example of cooling scenario profiles. Top: with spectral index change as in Fig. \ref{fig14}. Bottom: adiabatic cooling.}
 \label{fig15}
\end{center}

\end{figure}

\begin{figure}[t]
\begin{center}  
\includegraphics[width=0.8\linewidth]{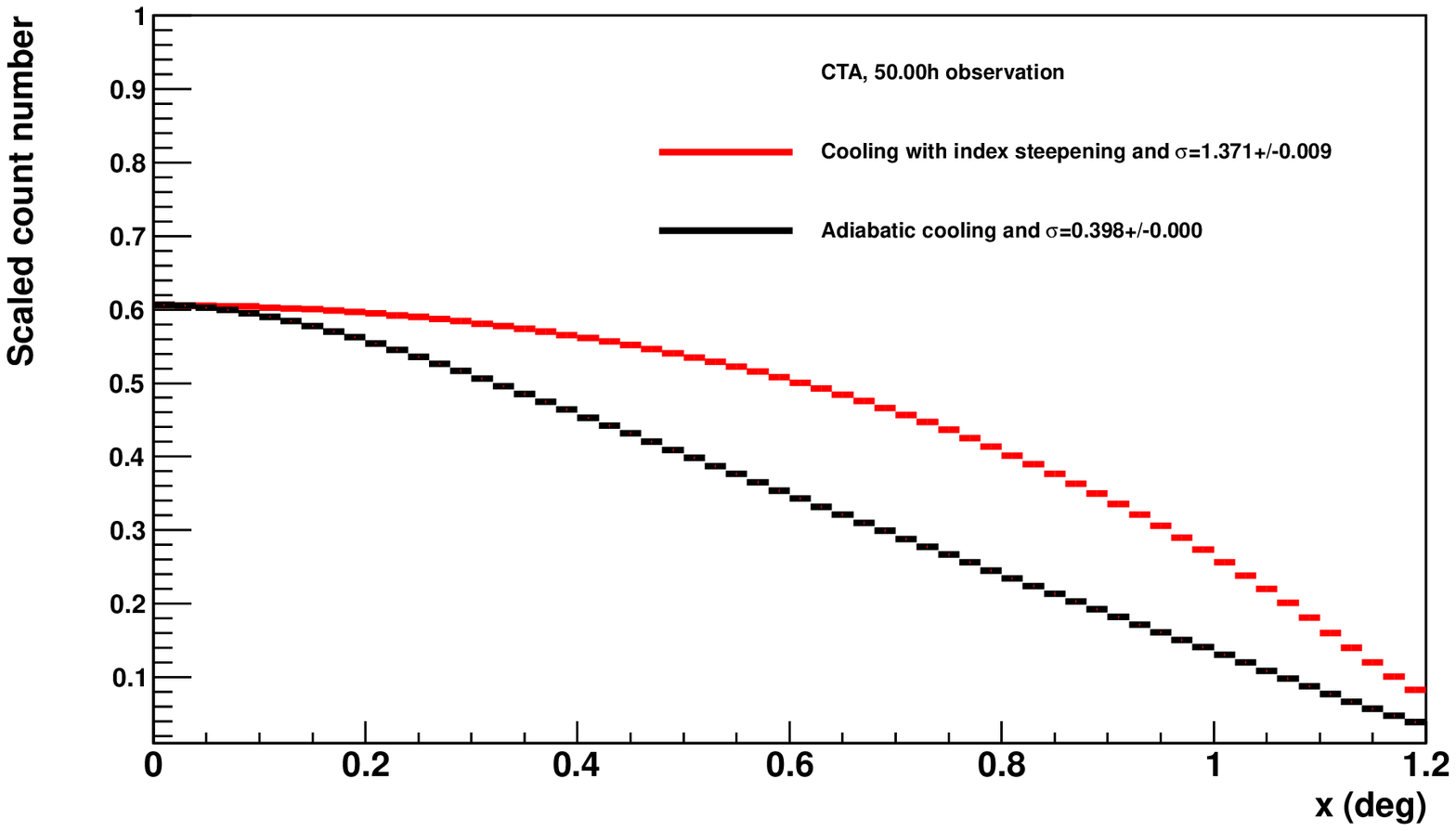}
\caption{The different fit extensions in an adiabatic or in a cooling for the same surface brightness distribution are shown. In black: The adiabatic case is simulated for a 0.4$^{\rm o}$ extended source. In red: A simulated case in which the index changes as in Fig. \ref{fig14} in a 1.4$^{\rm o}$ source.}
 \label{fig16}
\end{center}

\end{figure}

Maps can be derived for a more generalized case of cooling, with an exponentially decreasing surface brightness distribution and a steepening spectral index with increasing distance form the center of the object. The assumed distributions are shown in Fig. \ref{fig14}. 
Dividing the energy range in five logarithmically spaced energy bins, we obtain the profiles shown in Fig. \ref{fig15} (left). The profiles at lower energies tend to be more extended for the case of index steepening. In the case of constant index distribution, the extension of the source is not energy-dependent.
An adiabatic cooling scenario would lead to a less extended emission even for the same surface brightness distribution (see Fig. \ref{fig16}).
Both configuration E and B allow the reconstruction of the expected results. However, due to a better sensitivity, configuration E allows us to probe the energy dependent morphology to higher energies.


\subsubsection{Population Studies}

To understand the feasibility of population studies, including the completeness
and uniformity of the sample in the Galaxy, we have selected three PWNe detected in VHE gamma-rays, namely SNR\,G21.5--0.5, HESS J1356--645 and Kes~75, which
have relatively low 0.3--30 TeV luminosities (of the order of 10$^{34}$ erg s$^{-1}$) compared to the other VHE PWNe. Monte--Carlo simulations of these three nebulae with CTA have been performed, according to their spectral and morphological properties as measured with H.E.S.S., for different array configurations (CTA\_B, CTA\_D and CTA\_I\footnote{B (resp. D) is optimized for low-energy (resp. high-energy) events, while I is the balanced configuration which provides a better performance over the whole energy range}), assuming an observation time of 20~h with a mean zenith angle of 20$^{\circ}$. By varying the PWN distances from 1 to 20~kpc, we have calculated the so-called horizon of {\it detectability}, defined as the distance at which the source has a peak significance of 5 $\sigma$ (see \citep{SNRCTA}). In order to convert these horizon estimates into numbers of detectable PWNe with CTA, we have modeled the Galactic PSR/PWN distribution according to the \cite{vallee08} logarithmic spiral arms, together with the \cite{case98} galactocentric distribution, and an arm dispersion as a function of the galactocentric radius following the Galactic dust model of \cite{drimmel01}. Any potential displacement from pulsar birth place due to the kick velocity
has been ignored.

\begin{figure*}[t]
\centering
\includegraphics[width=0.5\textwidth]{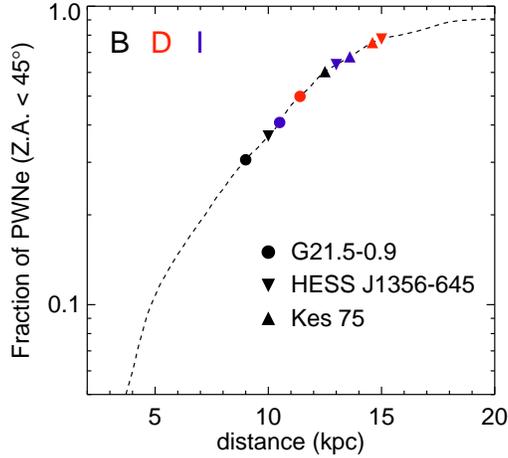}
\caption{Fraction of visible PWNe as a function of the distance to the Sun. For each source and each 
         CTA configuration considered here, filled and open symbols give 
         respectively the horizons of detectability and resolvability, as 
         defined in the text. 
}
\label{fig17}
\end{figure*}

The resulting source distribution is shown in Figure 
\ref{fig17}, together with the fraction of visible\footnote{The visibility is 
defined here as the fraction of the sky seen by an observatory located at the 
same latitude as the H.E.S.S. site, at zenith angles $\leq$45$^{\rm o}$.} PWNe 
as a function of the distance to the Sun.

A large fraction f$_{{\rm PWN}}$($\sim$ 0.4--0.8) of SNR\,G21.5--, 
HESS\,J1356-- and Kes75--like nebulae should be detectable with configuration D and I. de Jager et al, 2009 have estimated the lifetime of 
TeV-emitting leptons in such sources to be $\sim$40 kyr (for B=3$\mu$G, 
similar to what has been found in several PWNe such as Vela~X). This gives a total number of detectable G21.5--0.5-, HESSJ1356--645-- and Kes75--like PWNe with CTA of N$_{PWN}$ $\sim$ 800 f$_{PWN}$ ($\tau_{PWN}$/40kyr) ($\nu_{PWN}$/2), where $\nu_{PWN}$ is the Galactic VHE-emitting-PWN rate (in units of number of sources per century). This implies that $\sim$ (300-600) PWNe should be detected with CTA (for configurations CTA--I and CTA--D). These estimates should be taken with care, since we have implicitly assumed that the three considered PWNe are representative of the whole population of VHE nebulae, i.e. we have not accounted for any time evolution of the PWN sizes and luminosities, which both depend on many parameters.

\subsection{Pulsars}

\subsubsection{The Crab Pulsar}
The Crab pulsar is the only pulsar detected 
so far at VHE energies with Cherenkov
telescopes. In 2008, MAGIC announced the detection of the Crab pulsation above
25 GeV making use of an novel trigger system \cite{magic_science}. 
Contrary to what one could expect from EGRET \cite{Egret_Crab}, AGILE
\cite{Agile_Crab} (and later on Fermi \cite{Fermi_Crab}) 
observations, the characteristic double-peaked structure of the Crab light-curve was still visible above 25\,GeV. Recently,
VERITAS \cite{veritas} and MAGIC \cite{mono,stereo} have 
reported the detection
of the Crab pulsar at even higher 
energies, providing measurements of the pulsed
spectrum which in both cases is well described by a power-law, ruling out
in consequence 
an exponential cutoff shape of the spectrum proposed by Fermi LAT. 
VERITAS spectrum for the total pulsed emission (P1+P2) from 100\,GeV up
to 400\,GeV follows a power-law with photon index
around $3.8\pm0.5$. In turn, MAGIC has been able to measure the spectra of each
peak separately in a broader energy range, from 50 up to 400\,GeV. The spectra of
both peaks are compatible with power-law functions of photon indices $4.0\pm0.8$ and
$3.42\pm0.26$ for P1 and P2, respectively. 

\begin{figure}[ht]
  \centering
  \includegraphics[width=0.49\linewidth]{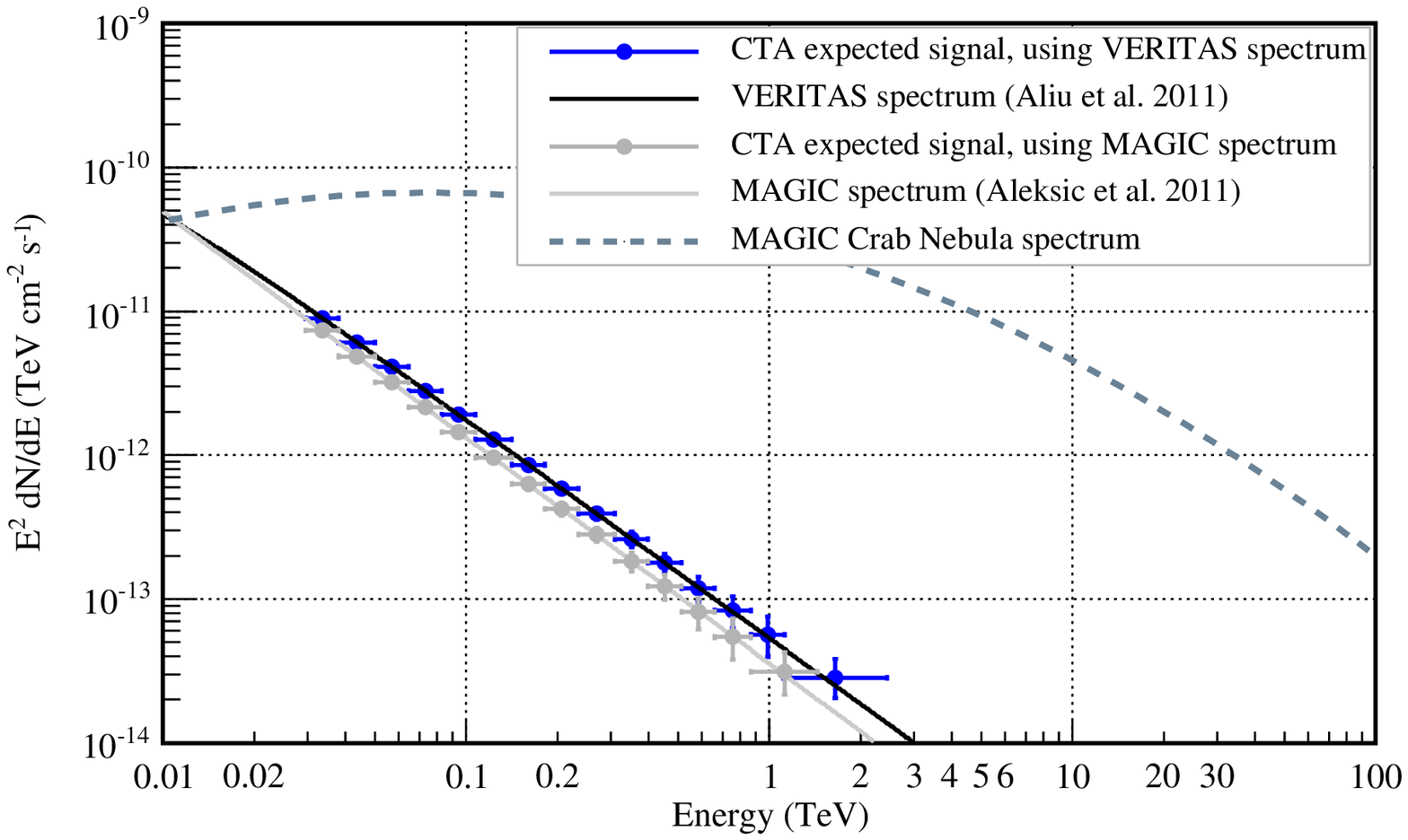}
  \includegraphics[width=0.49\linewidth]{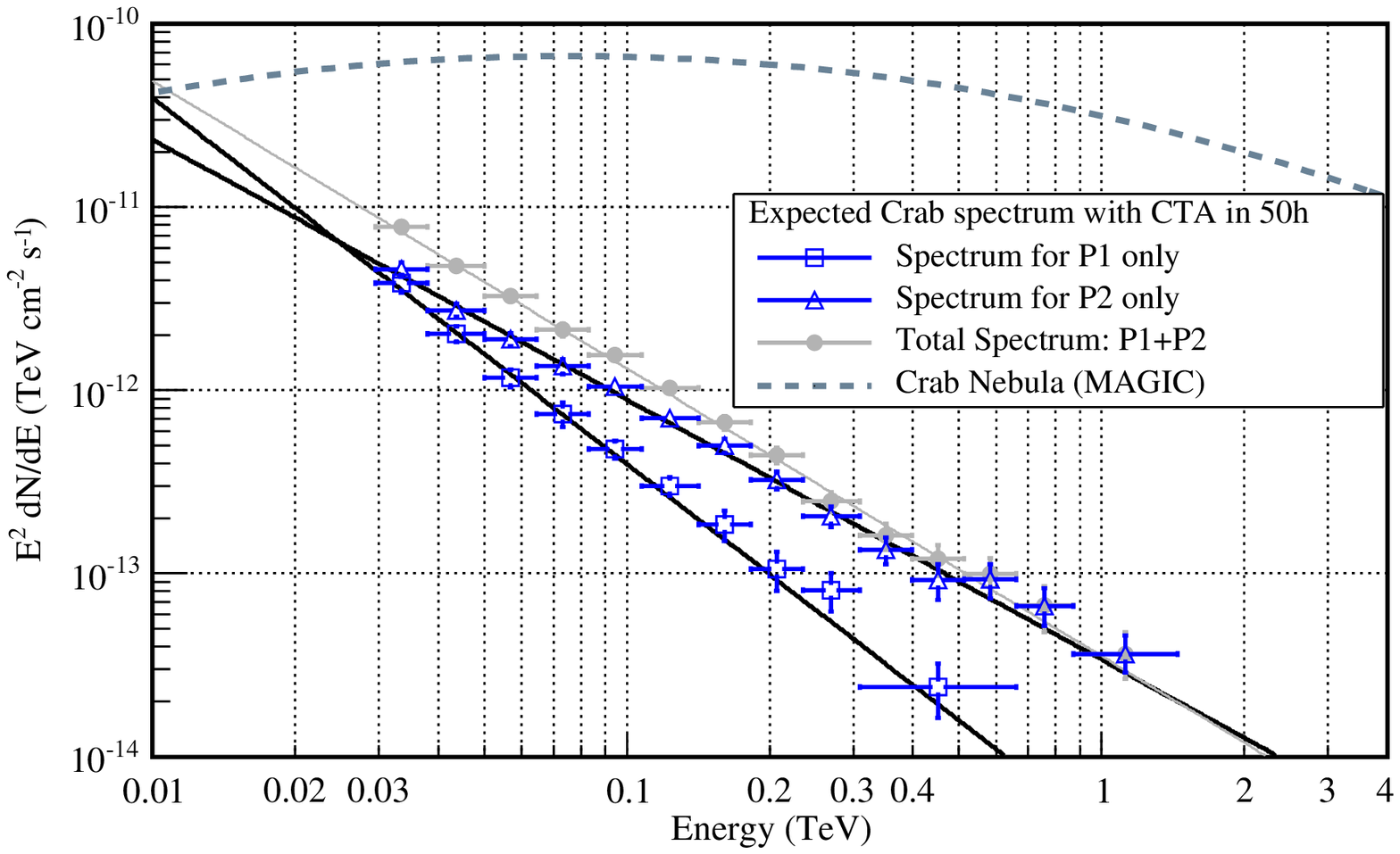}
  \caption{Left a): Simulated spectrum of the Crab Pulsar for 50h with
    CTA configuration B. The input models correspond to the joint fit (blue
   line)  of VERITAS and Fermi-LAT data used in
    \cite{veritas}, and the MAGIC power-law fit (grey line) from \cite{stereo}. 
    Right b): Simulated spectra of each Crab pulsar peak for 50h with CTA
    configuration B, using MAGIC power-law fits given in \cite{stereo}.
    The Crab Nebula spectrum from MAGIC is shown in both figures.} 
  \label{crab_fig} 
\end{figure}

In order to estimate the CTA potential for detecting the Crab pulsar, we have
generated a full simulated spectrum for a 50h observation with CTA
 configurations B, E and C using {\it CTAmacrosv6}. 
In the case of pulsar observations, the signal is expected only at certain
pulsar rotational phases, coinciding with the peaks seen in the pulsar pulse
profiles. One can then make use of the arrival
time of the photons to further discriminate between $\gamma$-ray-like events and
background (both hadronic and nebular).
The Crab light curve at VHE energies is characterized by two narrow peaks,
extending no more that a $10\%$ of the rotational period
\cite{veritas,stereo}. Therefore we have assumed a $90\%$ background reduction
coming from a proper timing analysis of the recorded events. 
Figure \ref{crab_fig} a) shows the results for CTA configuration B of
simulating the joint 
fit of the VERITAS and Fermi-LAT data reported in \cite{veritas}, as well
as the MAGIC spectrum for the total pulsed emission 
($P_{1M}$+$P_{2M}$) from \cite{stereo}. It 
is clear from the figure that CTA should be able to
collect sufficient photons to reveal the extent of the Crab pulsed emission in
energy out to at least 1\,TeV.  In fact, the bare detection  
of the pulsations would take less than one hour. Figure \ref{crab_fig} b)
shows the expected CTA spectra for both Crab peaks using MAGIC
measurements. Emission of P1 will vanish at energies above $\approx$500\,GeV. 
To quantify the goodness of the reconstruction, the simulated spectral
points were fitted to a power-law function. The results
of the fit and its uncertainties are listed in Table \ref{fits_table} for different CTA
 configurations. Configurations B and E
reconstruct the original emission accurately. Configurations
without large-size telescopes (LSTs), such as C, 
produces significantly higher errors in the fitting
parameters.

\begin{table}[ht]
\caption{Fit parameters with their uncertainties to a power-law function
  ($dN/dE=I_0 \times (E/100~GeV)^{-\Gamma}$)
for configurations B, E and C. } 
\resizebox{\textwidth}{!}{
\scriptsize
\begin{tabular}{ccccccc}
  \hline                      
  Configuration & \multicolumn{3}{c}{$I_{0} \times 10^{-11} (TeV^{-1} cm^{-2}
    s^{-1})$} & \multicolumn{3}{c}{Spectral Index} \\
\hline
    & (P1 + P2) & P1 & P2 & (P1 + P2) & P1 & P2 \\
\hline
  B & $13.62 \pm 0.03$ & $4.34 \pm 0.21$ & $9.10 \pm 0.24$ & $3.57 \pm 0.03$ & $3.89 \pm 0.07$ & $3.40 \pm 0.04$ \\
  E & $13.61 \pm 0.04$ & $4.51 \pm 0.26$ & $9.29 \pm 0.30$ & $3.58 \pm 0.04$ & $3.84 \pm 0.10$ & $3.40 \pm 0.04$ \\
  C & $14.95 \pm 1.10$ & $4.10 \pm 1.54$ & $9.11 \pm 0.75$ & $3.71 \pm 0.08$ &
  $3.72 \pm 0.85$ & $3.44 \pm 0.07$ \\
\hline

\end{tabular}
}
\label{fits_table}
\end{table}

%
\subsubsection{Prospects for new pulsar detections at VHE energies}

\begin{figure}[ht]
  \centering
 \includegraphics[width=0.8\linewidth]{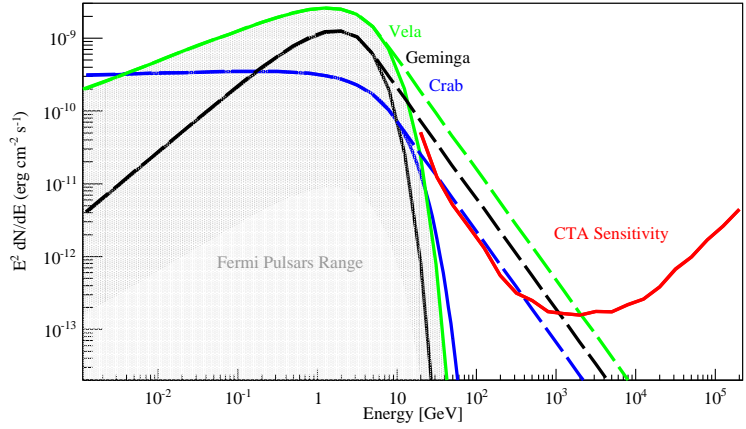}
  \caption{Pulsar spectra fits with power-law and exponential
    cutoff obtained by Fermi \cite{2fgl}  are compared to the CTA sensitivity curve
    for configuration B in 50 h.
    The fits for Vela, Crab and Geminga are indicated explicitly with
    continuos lines (red, blue and black, respectively). The shaded area contains
    the fits for the remaining 43 pulsars. 
     An extension of the Fermi pulsars' spectra to the CTA energy range by adding
     a power-law tail is shown with long-dashed lines for Vela, Crab and Geminga.
     The photon index $\beta$ of the power-law tail is taken equal to $3.52$
     (see text for explanation).}
  \label{spaghetti}
\end{figure}

With exponential cutoff energies between 0.7\,GeV and 7.7\,GeV
according to the 1st Fermi LAT Catalog of Gamma-ray Pulsars \cite{Fermi_pulsars}, the
current detection prospects 
with CTA for additional  {\it Fermi} pulsars appear dim at first.
Not even the Crab pulsar would be detected at VHE energies
under this scenario. To illustrate this
further,
Figure \ref{spaghetti} shows the
spectral fits (power-law with exponential cutoff) of the
Fermi pulsars taken from \cite{2fgl}, in comparison with the standard CTA
sensitivity curve for configuration B in 50 h. 
The fits of Vela, the Crab pulsar and Geminga are indicated explicitly
while 
the shaded area contains
the fits for the remaining 43 pulsars.

The Crab
detection by MAGIC and VERITAS discussed above, along with the fact that at the
highest energies measured by Fermi some pulsars already seem to deviate from
a simple exponential fit, led us to investigate 
what would happen if all the {\it Fermi}
pulsars were to present a power-law tail like
the one seen in Crab. 
While it is unclear that 
VHE emission is present in
every  pulsar, CTA could search for and place stringent constraints on
the putative pulsed emission above a few tens of GeV down to a sensitivity of 
0.001 c. u. 

\begin{figure}[ht]
  \centering
 \includegraphics[width=0.8\linewidth]{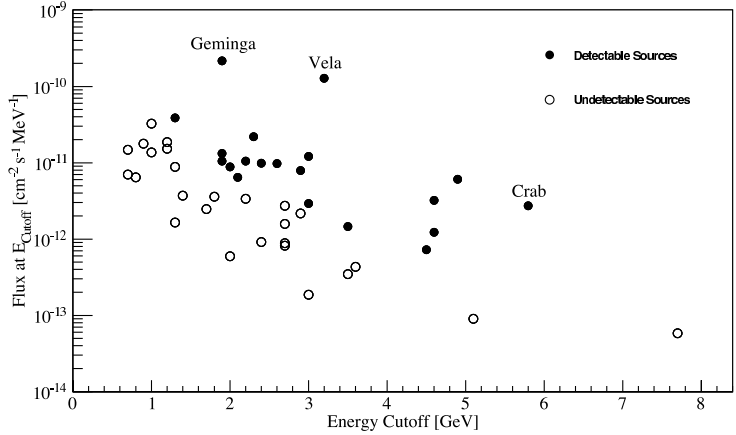}
  \caption{Pulsars detectable in 50 h with CTA configuration B under the assumption of Crab-like VHE tails
    as a function of their cutoff energy and flux.}
  \label{detect_fig}
\end{figure}

To test such {\sl Ansatz}, we have calculated how may
pulsars would be seen with CTA depending on the array configuration.
In practical terms, 
we have extended the spectra measured by Fermi above their cutoff energy
with a power-law tail that assumes 
the same spectral index as the one found for
the Crab by VERITAS, when a broken power-law is applied to 
fit both Fermi LAT and VERITAS detections, i.e $\beta = 3.52$. 
The final extrapolated 
spectral shapes for 3 out of 46 pulsars are shown in Figure \ref{spaghetti}).
With such hypothetical (except for the case of the Crab pulsar) additional power-law tails
all 46 pulsars were then considered as targets for 50 h observations with the CTA configurations: B, C and E.
We found that 20 pulsars would be detected with the configuration B and E; this number reduces to 12 
for configuration C. 
This indicates that
configurations B and E are better suited for pulsar 
studies than C.
Figure \ref{detect_fig} 
shows how the detectability with configuration B depends on the exponential cutoff energy value 
(as determined by Fermi LAT)
and the photon flux density at this energy.
In conclusion, it seems that
under the hypothesis of the
existence of the VHE Crab-like energy tails, 
a large fraction (up to $\sim 40$\% for configuration B and E) of the brightest Fermi pulsars may be detected with CTA.

\begin{figure}[ht]
  \centering  
  \includegraphics[width=0.8\linewidth]{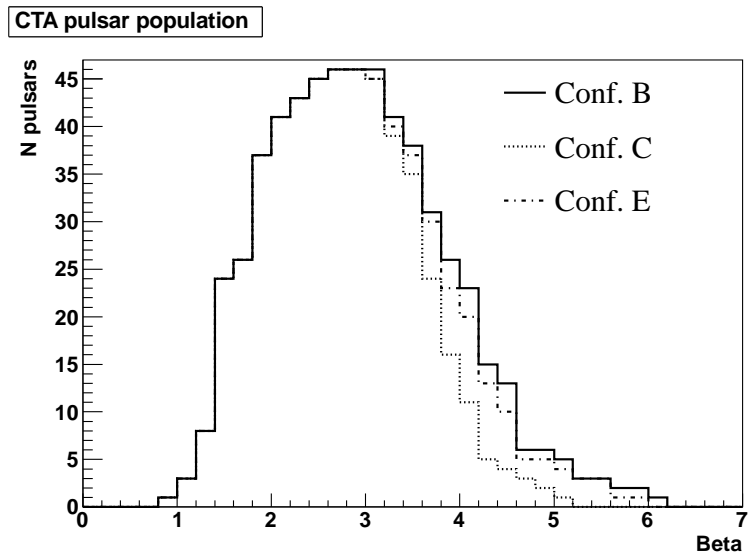}
 \caption{The detectability of 46 pulsars from \cite{Fermi_pulsars}
in 50 h under the assumption of power-law VHE tails 
    as a function of the slope $\beta$ (see text) for configurations B, C and E.}
 \label{raquel}
\end{figure}

On a second step, we can speculate on power-law tails in pulsar spectra 
with other slopes than $\beta = 3.52$. To test such cases
we used broken power-law spectral shapes in the form proposed by VERITAS \cite{veritas}.
The key parameters in this form are: E$_{\rm 0}$ - the break energy (usually, quite close to the exponential cutoff energy),
 $\alpha$ -the slope of the photon flux spectrum in the Fermi LAT range well below E$_{\rm 0}$, $\beta$ - the slope
of the photon flux spectrum in the tail, i.e. well above E$_{\rm 0}$ 
(note: we define $\alpha$ and $\beta$ with a sign opposite to
the original notation in \cite{veritas}). For all 46 pulsars
the values of E$_{\rm}$ and $\alpha$ were used based on \cite{Fermi_pulsars}.  
With the limit of 50 h observation time on one hand and the condition 
of the overall spectral shape to be convex
 $\beta > \alpha$ on the other hand, the detectability of these pulsars
in function of $\beta$ was studied for configurations B, C and E.
As expected, configuration B is the optimal
one for all the possible values of $\beta$; the second
best is configuration E and the worst one is C (see Figure \ref{raquel}).

For $\beta = 3.52$ the results presented in the previous paragraph
should be identical to those of Figure \ref{detect_fig}. This is not the case -
the latter results give higher percentage of detectable pulsars than the former ones.
The reason for this is that each analysis was performed by an independent group.
The two groups used somewhat different analytical functions to approximate
the HE-VHE pulsar spectra. Moreover, their treatment
of background effects, of pulsed fraction as a function of energy for
each pulsar (timing plays an important role in improving the sensitivity) etc., was not identical.
However, in both cases the detectability of pulsars with putative
VHE tails is quite high - no less than $\sim 40\%$. 
Needless to say, there is little hope that all gamma-ray pulsars will cooperate in
the way described above. However, some theoretical models of young and energetic pulsars as well as
old millisecond pulsars speak in favor of pulsed spectral components located in the VHE domain.
CTA will be the only
facility in near future capable of solving this problem, at least to some extent. 

\subsubsection{Globular Clusters}

The two main candidates for detections of unpulsed VHE emission from
globular clusters are 47 Tuc and Ter 5 due to their large
population of MSPs. 

\begin{figure}
 \centering
 \includegraphics[width=\textwidth]{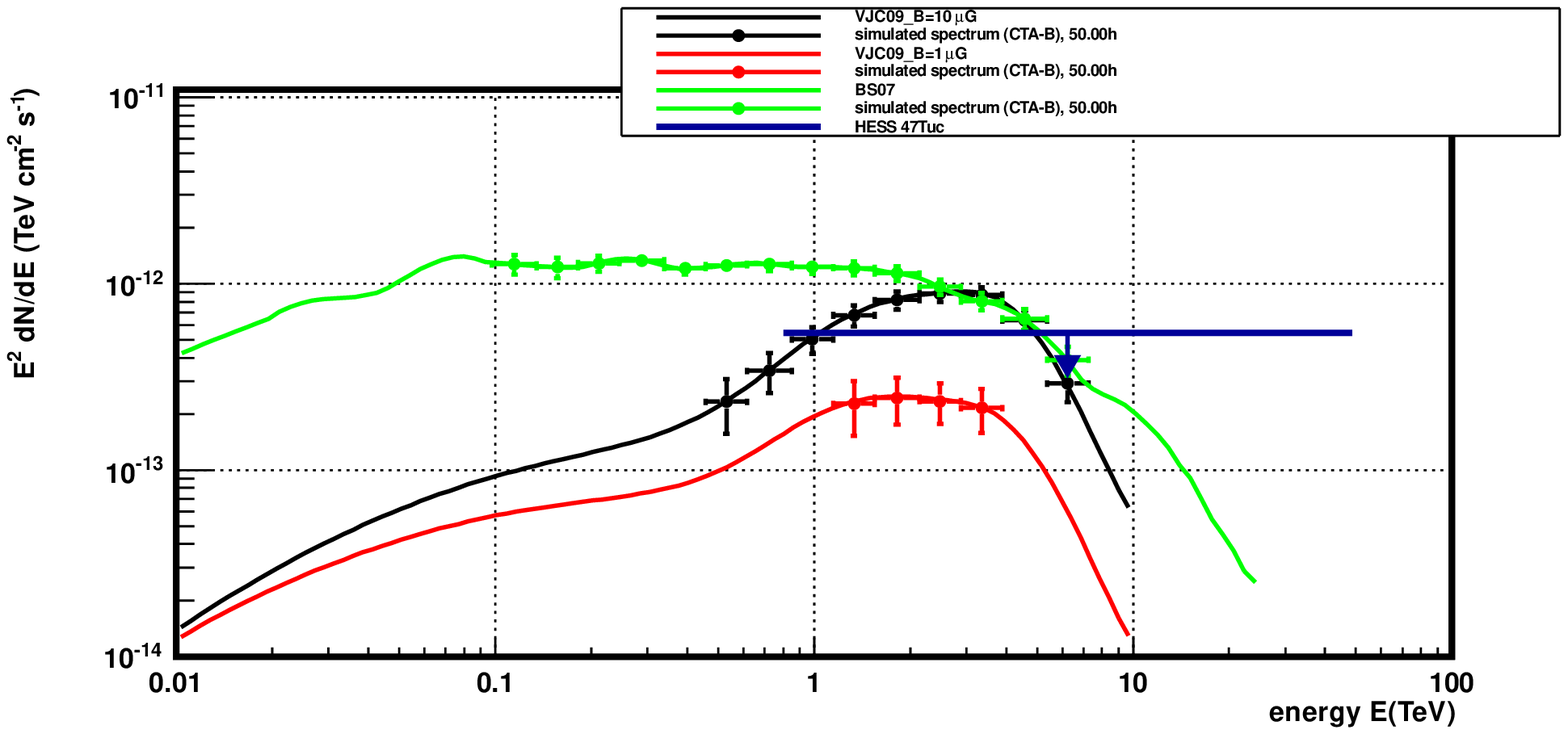} 
  \includegraphics[width=\textwidth]{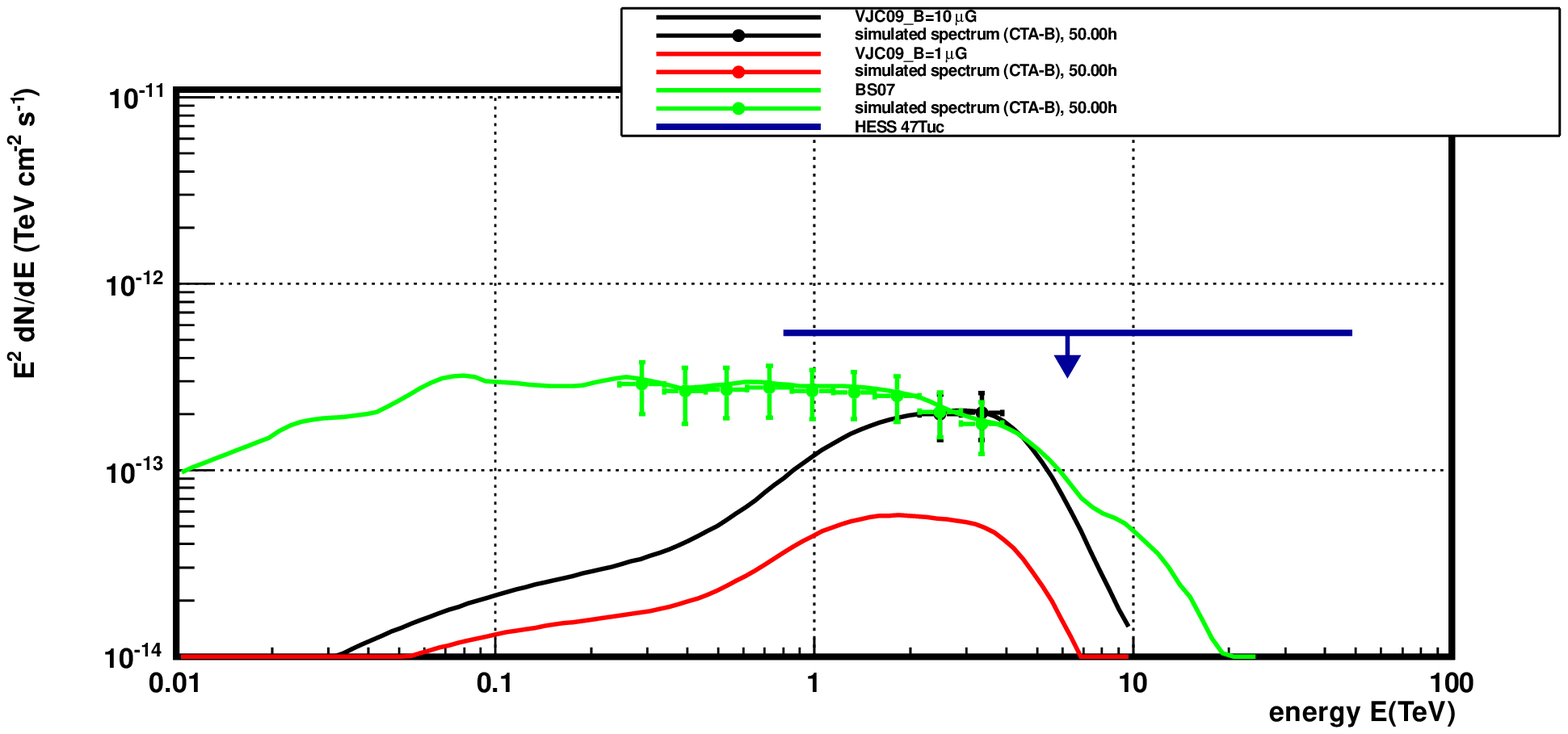} 
 \caption{The expected CTA response in configuration B for the models of 47 Tuc
by \cite{bs} (green line) and \cite{ven} (black line for high magnetic field in the cluster,
red line for low magnetic field). The H.E.S.S. upper limit is shown in blue. Top a) the results for 
assumed 100 MSPs in the cluster. Bottom b) rescaled to the number of 23 (as known) MSPs.}
 \label{47tuc}
\end{figure}

\begin{figure}
 \centering
 \includegraphics[width=\textwidth]{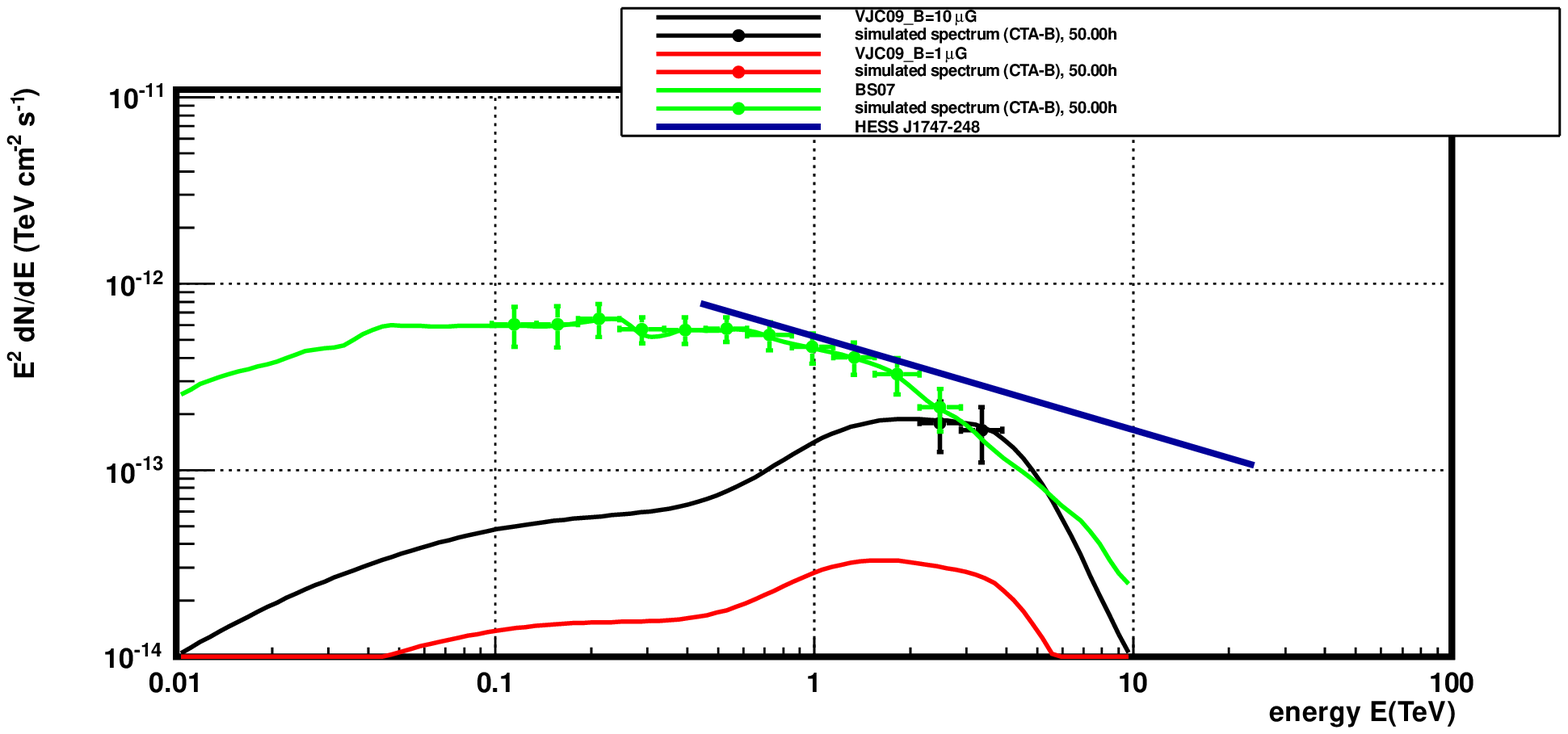} 
  \includegraphics[width=\textwidth]{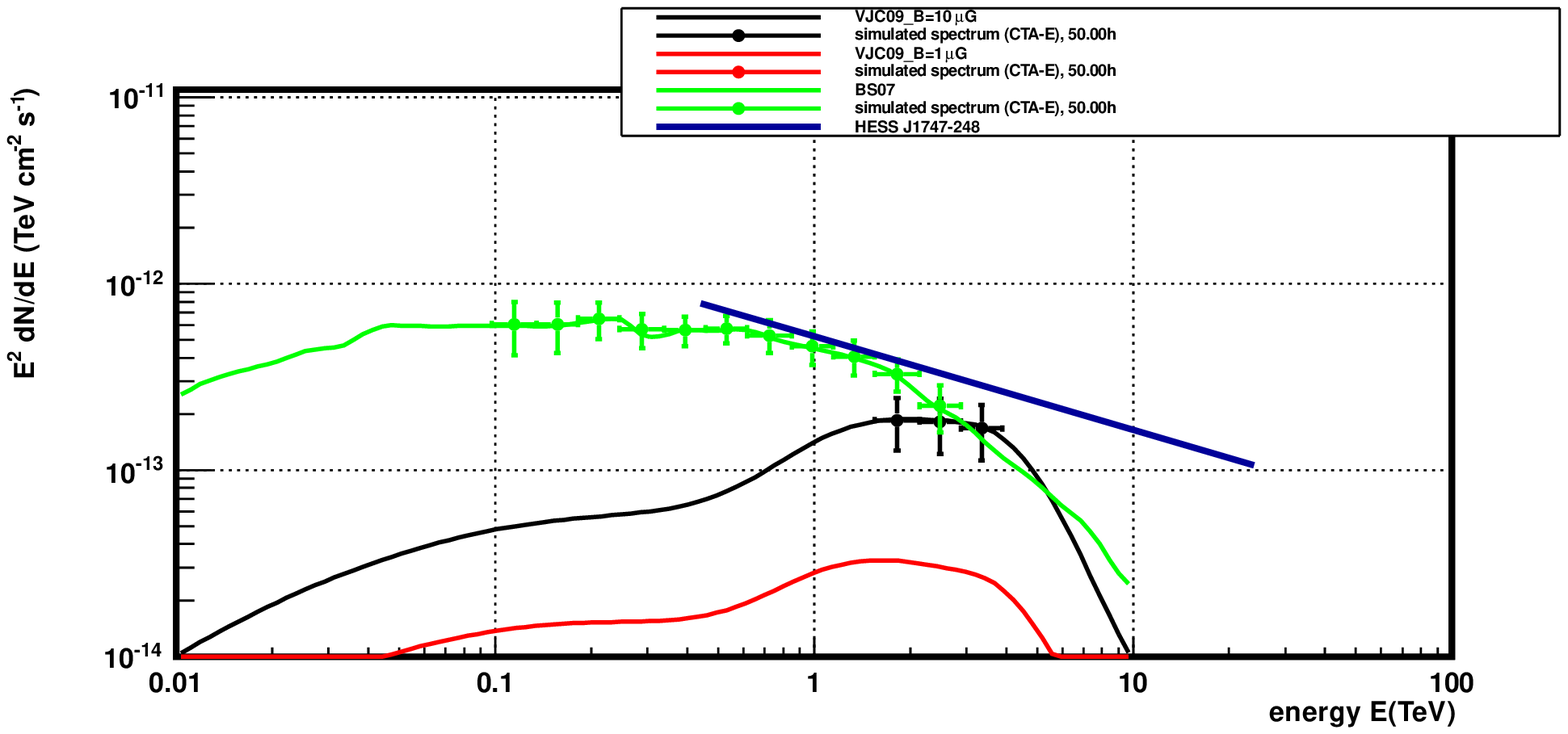} 
 \caption{The expected CTA response for the models of Ter 5
by \cite{bs} (green line) and \cite{ven}.  (black line for high magnetic field in the cluster,
red line for low magnetic field). The level of HESS\,J1747--248 is shown in blue. Configuration B is shown on the top a) while the E one is shown on the bottom b)}
 \label{ter5}
\end{figure}

To assess the detectability of these two clusters, we evaluate only the CTA response using configuration E and B. Indeed, due to the expected sharp drop of the emission at energies of 10 TeV, we assumed that a configuration optimized for the high energies would not bring any improvement.
Simulations using prediction from \cite{bs} and \cite{ven} (calculated for a population of 100 MSPs) are shown in Fig. \ref{47tuc} and \ref{ter5}. From the figure it can be derived that the detection of 47~Tuc is expected in all the explored models with observation times up to 50 hours. Since the VHE spectrum is linearly dependent on the number of MSPs in the system the model relative to 47~Tuc is also scaled by a factor of 0.23 (relative to considering only the 23 detected MSPs in the system). For this case, CTA will also be able to detect it even when high magnetic field values are considered.
For the second candidate, Ter~5 the fluxes predicted are lower. However, detection can still be achieved with a moderately long observation time.
For comparison, we show also the best fit spectrum from the 90 h observation of HESS\,J1747--248 (positionally coincident with Ter~5). Different models were re-scaled and discussed in the detection paper. Observations with CTA will provide a detailed study of the spectrum and will allow us to associate it or not with the cluster itself (see Fig. \ref{47tuc} and \ref{ter5} for the simulations).
Only the response from configuration E and B was investigated.
The results for the two configurations considered are comparable. Indeed, the low energies for which
configuration B is optimized are not crucial for the distinction of the
models investigated here.

\section{Outlook}

The wide range of phenomenological research comprised on the studies of PWNe and Pulsars requires diverse specifications. 

For PWN, in general, the simulations described above tend to favor a balance configuration (either D or E) in comparison with the configurations B and C, optimized for low and high energies, respectively. In particular, the horizon of detectability optimizes for configuration D and/or E as well as the sensitivity to cooling effects, given the larger range of energy covered with good sensitivity. Moreover, although the detailed morphological and spectral investigations show an improvement when using configurations which are optimized for low and high energies for these particular cases, an intermedium solution seems the most sensible compromise for PWNe. 

The prospects for high-quality pulsar studies with CTA should be considered with caution; the Fermi LAT results are quite meaningful.
A sizable progress in studies of individual pulsars will be possible with the configurations optimized for very low energy range, especially below 50 GeV. None of the planned configurations will be sensitive enough to deal with pulsed radiation of Fermi LAT-type energy cutoffs.
However, the discovery by VERITAS and MAGIC of pulsed radiation
from the Crab pulsar well above the LAT-inferred exponential cutoff
energy came as a surprise and yet another example supporting a statement
by Trevor Weekes  twelve years ago that
''VHE astronomy is still observation-driven and that theoretical VHE
astrophysics still lags as a predictive discipline" \cite{weekes99}.

If other LAT pulsars are similar to the Crab pulsar and contain a power-law type turnover
then configurations B or E may detect a good fraction of these pulsars.
More promising results with B or E configurations are expected
in future studies of globular clusters. The actual performance
of B and E configurations
will depend strongly on (still unknown) population of MSPs in GCs as well as
on physical properties of ambient fields in the GCs like the strength of magnetic field 
and the field of ambient soft photons. In principle CTA may be able to study
both spectral shapes as well as angular extension of the VHE radiation from GCs.

\noindent
\\
\\
\\
\\
\\
\\
\\
\\
\\
\\
B.R. acknowledges support through the grants MNiSW N203387737 and NCBiR ERA-NET-ASPERA/01/10.
J. A. B., J. L. C., T. H., M. L. and N. M. acknowledge the support of the Spanish MICINN under project code 
FPA2010-22056-C06-06. E. de O. W., M. L. and N. M. gratefully acknowledge support from the Spanish MICINN through a
 Ram\'on y Cajal fellowship.
D.H., G.P and D.F.T. acknowledge support from the Ministry of Science and the
Generalitat de Catalunya, through the grants AYA2009-07391 and
SGR2009-811, as well as by ASPERA-EU through grant EUI-2009-04072 and the Formosa Program TW2010005.

\end{document}